\documentclass[aps,floats]{revtex4-2}
\usepackage{amsmath,amssymb}
\usepackage{graphicx,epsfig}
\usepackage[greek,english]{babel}
\usepackage{bbold}

\DeclareMathOperator{\arctanh}{arctanh}
 
\begin{document}
\bibliographystyle {plain}

\pdfoutput=1
\def\oppropto{\mathop{\propto}} 
\def\opsimeq{\mathop{\simeq}}
\def\opoverderline{\mathop{\overline}}
\def\operarrow{\mathop{\longrightarrow}}
\def\opsim{\mathop{\sim}}

\def\opmin{\mathop{\min}} 
\def\opmax{\mathop{\max}} 
\def\oplim{\mathop{\lim}}

\title{ Markov spin models for image generation : \\
explicit large deviations with respect to the number of pixels } 


\author{C\'ecile Monthus}
\affiliation{Universit\'e Paris-Saclay, CNRS, CEA, Institut de Physique Th\'eorique, 91191 Gif-sur-Yvette, France}


\begin{abstract}
For the discrete-time or the continuous-time Markov spin models for image generation when each pixel $n=1,..,N$ can take only two values $S_n=\pm 1$, the finite-time forward propagator depends on the initial and on the final configurations of the $N$ spins only via a single global variable, namely the extensive overlap that counts the number of spins that have the same value or not in the two configurations. The joint probability distribution of the overlap and of the magnetization during the forward noising dynamics can be written for any finite number $N$ of pixels and in the limit $N \to + \infty$ to extract the large deviations properties. The consequences for the backward reconstructive dynamics are then analyzed for various initial conditions, namely (i) a single image (ii) a mixture of two images (iii) when the initial condition corresponds to the Curie-Weiss mean-field ferromagnetic model in the microcanonical ensemble, as a simple analog of the manifold-hypothesis concerning continuous generative diffusion models.

\end{abstract}

\maketitle


\section{ Introduction  }

Besides their algorithmic and practical interests,
the generative Markov models (see the review \cite{GeneReview} and references therein)
are also fascinating in relation with other areas of statistical physics and of stochastic processes,
as stressed from various perspectives in the recent works \cite{gene_noneq,Biroli_largedim,Ambrogioni1,Ambrogioni2,Ambrogioni3,Biroli_dynamical,emergence,pathintegral,GeneNonEq,wasserstein,Ambrogioni4,Ambrogioni5,cumulants,garrahan,c_generativePropagator}.
The forward dynamics transforming incrementally non-trivial images into pure noise
is an example of the transient non-equilibrium relaxation
towards equilibrium when starting from arbitrary initial conditions, 
that can be analyzed via the spectral decomposition of the forward propagator \cite{c_generativePropagator}.
The backward reconstructive dynamics 
is based on the mathematical possibility of reversing the time-arrow in Markov processes
that can be implemented on computers, even it would never naturally occur in real life.
Its ability to 'create order out of noise' might be surprising at first, but can be understood if one thinks of the artificial 
time-dependent backward generator as some kind of 'stochastic Maxwell demon'  
that is sufficiently well informed to drive the system from high entropy towards low entropy.
The analysis of this backward reconstructive dynamics
actually raises a lot of interesting issues for statistical physicists, 
in particular in relation with spontaneous symmetry breaking, memorization and glass transitions (see \cite{Biroli_largedim,Ambrogioni1,Ambrogioni2,Ambrogioni3,Biroli_dynamical,emergence,Ambrogioni4,Ambrogioni5}
and references therein).
So while the practical applications of generative Markov models have recently exploded with very impressive achievements,
it seems nevertheless useful to develop a better theoretical understanding 
from the point of view of statistical physics when the number $N$ of pixels becomes large $N \to + \infty$.

In statistical physics, whenever there is a big parameter,
the theory of large deviations (see the reviews \cite{oono,ellis,review_touchette} and references therein)
is used either implicitly (just like Mr Jourdain speaking in prose without knowing it)
or explicitly. The explicit use of the language of large deviations has recently played a major role 
in the unification between the equilibrium part and the non-equilibrium part of statistical physics,
 and has led to a much deeper understanding of non-equilibrium stochastic processes
(see the reviews with various perspectives \cite{derrida-lecture,harris_Schu,searles,harris,mft,sollich_review,lazarescu_companion,lazarescu_generic,jack_review}, 
the PhD Theses \cite{fortelle_thesis,vivien_thesis,chetrite_thesis,wynants_thesis,chabane_thesis,duBuisson_thesis} 
 and the Habilitation Thesis \cite{chetrite_HDR}).
 In particular, many results have been obtained in two important asymptotic regimes, namely
on one hand for time-averaged observables over a single stochastic trajectory observed during a long time-window $T \to + \infty$
 \cite{peliti,derrida-lecture,sollich_review,lazarescu_companion,lazarescu_generic,jack_review,vivien_thesis,lecomte_chaotic,lecomte_thermo,lecomte_formalism,lecomte_glass,kristina1,kristina2,JackSollich2010,simon1,simon2,tailleur,simon3,Gunter1,Gunter2,Gunter3,Gunter4,chetrite_canonical,chetrite_conditioned,chetrite_optimal,chetrite_HDR,touchette_circle,touchette_langevin,touchette_occ,touchette_occupation,garrahan_lecture,Vivo,c_ring,c_detailed,chemical,derrida-conditioned,derrida-ring,bertin-conditioned,touchette-reflected,touchette-reflectedbis,c_lyapunov,previousquantum2.5doob,quantum2.5doob,quantum2.5dooblong,c_ruelle,lapolla,c_east,chabane,us_gyrator,duBuisson_gyrator,c_largedevpearson,sollich_MFtrap}, 
 and on the other hand for time-dependent properties in systems involving a large number $N \to + \infty$ of degrees of freedom, 
 either via the Macroscopic Fluctuation Theory for driven diffusive systems (see the review \cite{mft} and references therein)
 or via the formulation in terms of Poisson processes for discrete configuration spaces \cite{maes_onandbeyond,c_interactions,c_open,chemical,chabane}.
 In the present work concerning generative Markov models, 
 our goal will be similarly to analyze the time-dependent properties
 of the forward noising dynamics and of the backward denoising
  dynamics when the number $N$ of pixels is large $N \to + \infty$.
 
The field of generative Markov models is dominated by the formulation in terms of diffusion processes,
but the case of Markov chains on discrete-state spaces has also been considered \cite{garrahan,c_generativePropagator,Discrete_stuctured,Discrete_denoising,Discrete_markov,Discrete_score,Discrete_guidance}. Since computers only work with discrete data, the diffusion models actually 
need in practice to be discretized both in configuration-space and in time. As a consequence, 
it seems useful to develop also the field of discrete-state discrete-time generative Markov models
 in order to have theoretical descriptions in direct correspondence with the computer implementations.
The simplest discrete-state framework is when each pixel $n$ can take only two values $S_n=\pm 1$ 
corresponding to the familiar spins of statistical physics and to the black-and-white images of old times; 
the case of colored images where each pixel can be in 
$C>2$ different colors would be of course more realistic for practical applications 
and is also standard in statistical physics from the Potts and clock models, but will not be discussed
 in the present paper in order to simplify the notations and the discussions.
 For these generative Markov models involving spins $S_n=\pm 1$,
the simplest continuous-time single-spin-flip model has been recently analyzed
 in \cite{garrahan,c_generativePropagator},
but one can alternatively consider its discrete-time counterpart as will be discussed in detail in section \ref{sec_general}.
Since these two dynamics actually share exactly the same finite-time forward propagator (see the more detailed discussion in subsection \ref{subsec_linkContoinuous}), one obtains the same conclusions for all properties concerning finite-time-intervals. 
The present work is based on the following key observation: this finite-time forward propagator 
depends on the initial configuration of the $N$ spins
and on the final configuration of the $N$ spins only via a single global variable with a simple physical meaning,
namely the extensive overlap $Q$ that counts the number of spins that have the same value or not in the two configurations. 
It is thus possible to make simple explicit calculations either for a finite number $N$ of pixels
or in the limit $N \to + \infty$ to extract the large deviations properties.

The paper is organized as follows.
In section \ref{sec_general}, we describe the general properties of 
discrete-time generative Markov spin models, as well as 
the specific form of the forward propagator in terms of the overlap between the initial and the final configurations.
In section \ref{sec_jointMQ}, the joint probability distribution of the extensive overlap $Q$ and of the extensive magnetization $M$ during the forward noising dynamics is written for any finite number $N$ of pixels and in the limit $N \to + \infty$ to extract the large deviations properties for the intensive overlap $q=\frac{Q}{N}$ and for the intensive magnetization $m=\frac{M}{N}$.
 The consequences for the backward-generative dynamics
 are then analyzed for three cases of initial conditions, namely 
 for a single initial image in section \ref{sec_single},
 for a mixture of two initial images in section \ref{sec_twoimages},
 and finally for the Curie-Weiss mean-field ferromagnetic model in the microcanonical ensemble in section \ref{sec_curie}.
Our conclusions are summarized in section \ref{sec_conclusions}, while
some technical details of the large deviations calculations are relegated to Appendices.


\section{ Discrete-time Markov spin models for image generation  }

\label{sec_general}

In this section, we introduce the notations that will be useful in the whole paper
to analyze discrete-time Markov spin models for image generation with $N$ pixels,
when each pixel $n=1,..,N$ has only two possible states $S_n=\pm 1$,
so that there are ${\cal N}=2^N $ possible images $\vec S =(S_1,..,S_N)$.

\subsection{ General properties of the forward noising dynamics and of the backward reconstructive dynamics}

The goal of this subsection is to give a short self-contained summary of the main ideas 
of Markov generative models within the discrete-time formulation which is more pedagogical
to understand the subtleties of the backward-generative dynamics.

\subsubsection{ Probabilities of the forward trajectories $\vec S(0 \leq t \leq T) $ over the time window $t=0,1,,,,T-1,T$ }

When the initial condition $\vec S(t=0) $ is drawn with the probability $ P^{ini}(\vec S(0) )$,
the probability ${\cal P}^{Traj} [ \vec S(0 \leq t \leq T)]  $ of the whole forward trajectory $\vec S(0 \leq t \leq T)$
 during the time window $t=0,1,,,,T-1,T$ 
simply involves the product over the time $t$ of the matrix elements $\langle \vec S(t) \vert {\cal W} \vert \vec S(t-1) \rangle $
of the Markov matrix $ {\cal W} $ between the image $\vec S(t-1) $ at time $(t-1)$ and the image $\vec S(t) $ at time $t$
 \begin{eqnarray}
{\cal P}^{Traj} [ \vec S(0 \leq t \leq T)] && 
= P^{ini}(\vec S(0) ) \prod_{t=1}^T \langle \vec S(t) \vert {\cal W} \vert \vec S(t-1) \rangle 
\label{Ptrajforward}
\end{eqnarray}
The specific choice of $ {\cal W} $ will be discussed later in subsection \ref{subsec_choice},
while here we describe the general principles.

The probability of the forward trajectory of Eq. \ref{Ptrajforward}
can be expanded more explicitly as
 \begin{eqnarray}
 {\cal P}^{Traj} [ \vec S(0 \leq t \leq T)]  
= \langle \vec S(T) \vert {\cal W} \vert \vec S(T-1) \rangle 
\langle \vec S(T-1) \vert {\cal W} \vert \vec S(T-2) \rangle
 .... \langle \vec S(2) \vert {\cal W} \vert \vec S(1) \rangle
\langle \vec S(1) \vert {\cal W} \vert \vec S(0) \rangle
P^{ini}(\vec S(0) )  
\label{Ptrajforwardexpanded}
\end{eqnarray}
This trajectory probability is the analog of the Feynman path-integral integrand for continuous diffusion models.

The summation of the trajectory probability of Eq. \ref{Ptrajforwardexpanded}
over all the spins except the configuration $\vec S(t)$ at time $t$
produces the probability $P_t(\vec S(t)) $ to see the configuration $\vec S(t) $ at time $t$ 
when the initial condition is $P^{ini}$
 \begin{eqnarray}
P_t(\vec S(t))  && = \sum_{\vec S(0 \leq \tau \leq t-1)} \sum_{\vec S(t+1 \leq \tau' \leq T)}{\cal P}^{Traj} [ \vec S(0 \leq t \leq T)]
\nonumber \\
&&  = \sum_{\vec S(0)}\langle \vec S(t) \vert {\cal W}^t \vert \vec S(0) \rangle P^{ini}(\vec S(0) )
  = \langle \vec S(t) \vert {\cal W}^t \vert P^{ini} \rangle   
\label{Ptforwardinteg}
\end{eqnarray}
while the summation over all the spins except the spins $\vec S(t_1)$ and $\vec S(t_2) $ at the two times $t_1<t_2$
produces the joint distribution
 \begin{eqnarray}
P_{t_2,t_1}(\vec S(t_2); \vec S(t_1))  
&& = \sum_{\vec S(0 \leq \tau \leq t_1-1)} \sum_{\vec S(t_1+1 \leq \tau' \leq t_2-1)}
\sum_{\vec S(t_2+1 \leq \tau'' \leq T)}
{\cal P}^{Traj} [ \vec S(0 \leq t \leq T)]
\nonumber \\
&&   = \langle \vec S(t_2) \vert {\cal W}^{(t_2-t_1)} \vert \vec S(t_1) \rangle \langle \vec S(t_1) \vert {\cal W}^{t_1} \vert P^{ini} \rangle
\label{Ptforwardintegt1t2}
\end{eqnarray}


\subsubsection{ Time-dependent backward generator $\langle \vec S(t) \vert W^B_{t+1/2}\vert \vec S(t+1) \rangle $ 
that reproduces the probabilities of the forward trajectories}

The forward trajectory probability of Eq. \ref{Ptrajforwardexpanded}
can be reproduced via a backward Markov analysis 
starting with the final distribution $P_T(\vec S(T)) $ of Eq. \ref{Ptforwardinteg} 
 \begin{eqnarray}
&& {\cal P}^{Traj} [ \vec S(0 \leq t \leq T)] 
 = P_T(\vec S(T) )  \prod_{t=0}^{T-1} \langle \vec S(t) \vert {\cal W}^B_{t+1/2}\vert \vec S(t+1) \rangle
\nonumber \\
&&= \langle \vec S(0) \vert {\cal W}^B_{1/2} \vert \vec S(1) \rangle 
\langle \vec S(1) \vert {\cal W}^B_{3/2} \vert \vec S(2) \rangle
 .... \langle \vec S(T-2) \vert {\cal W}^B_{T-3/2} \vert \vec S(T-1) \rangle
\langle \vec S(T-1) \vert {\cal W}^B_{T-1/2} \vert \vec S(T) \rangle
P_T(\vec S(T) )  
\label{Ptrajbackward}
\end{eqnarray}
where the matrix elements $\langle \vec S(t) \vert {\cal W}^B_{t+1/2}\vert \vec S(t+1) \rangle $
of the time-dependent backward Markov generator ${\cal W}^B_{t+1/2} $ 
 \begin{eqnarray}
\langle \vec S(t) \vert {\cal W}^B_{t+1/2}\vert \vec S(t+1) \rangle  
&& \equiv \langle \vec S(t+1) \vert  {\cal W} \vert \vec S(t) \rangle 
\frac{  P_{t}(\vec S(t)) }{ P_{t+1}(\vec S(t+1)) }
= \langle \vec S(t+1) \vert  {\cal W} \vert \vec S(t) \rangle 
\frac{  \langle \vec S(t) \vert {\cal W}^t \vert P^{ini} \rangle }
{ \langle \vec S(t+1) \vert {\cal W}^{t+1} \vert P^{ini} \rangle }
\label{WBchain}
\end{eqnarray}
involves the reweighting of the forward matrix element $\langle \vec S(t+1) \vert  {\cal W} \vert \vec S(t) \rangle $
by the ratio $\frac{  \langle \vec S(t) \vert {\cal W}^t \vert P^{ini} \rangle }
{ \langle \vec S(t+1) \vert {\cal W}^{t+1} \vert P^{ini} \rangle } $ of two forward propagators starting at the initial condition $P^{ini} $.
The logarithm of this ratio is the discrete-space discrete-time analog of the score function $\left( - \vec \nabla \ln  p_t(\vec x) \right)
= \left( - \vec \nabla \ln   \langle \vec x \vert {\cal W}^t \vert p^{ini} \rangle \right) $ that appears as an effective force in the backward Langevin dynamics of generative diffusion models.

Since this time-dependent backward Markov generator ${\cal W}^B_{t+1/2} $
has been defined in order to reproduce the full trajectories probabilities of Eq. \ref{Ptrajbackward},
it will automatically reproduce all the marginal properties that can be obtained via summations over some variables
in Eq. \ref{Ptrajbackward}, with the following important consequence
 for the backward propagator over any finite-time interval.


\subsubsection{ Backward propagator ${\cal B}_{t_1,t_2}[ \vec S(t_1) \vert \vec S(t_2) ] $ 
over any finite-time interval $[t_1,t_2]$}

The backward propagator ${\cal B}_{t_1,t_2}[ \vec S(t_1) \vert \vec S(t_2) ]  $ between the configuration $\vec S(t_2) $ at time $t_2$ and the configuration $\vec S(t_1) $ at the smaller time $t_1<t_2$ 
 \begin{eqnarray}
{\cal B}_{t_1,t_2}[ \vec S(t_1) \vert \vec S(t_2) ] && \equiv \langle \vec S(t_1) \vert {\cal W}^B_{t_1+1/2} ... {\cal W}^B_{t_2-1/2} \vert \vec S(t_2) \rangle  
= \sum_{\vec S(t_1 <\tau<t_2) }  \prod_{t=t_1}^{t_2-1} \langle \vec S(t) \vert {\cal W}^B_{t+1/2}\vert \vec S(t+1) \rangle
\nonumber \\
&& = \sum_{\vec S(t_1 <\tau<t_2) }  \prod_{t=t_1}^{t_2-1}
\left[ \langle \vec S(t+1) \vert  {\cal W} \vert \vec S(t) \rangle \frac{  P_{t}(\vec S(t)) }
{ P_{t+1}(\vec S(t+1)) } \right]
\nonumber \\
&&= \langle \vec S(t_2) \vert  {\cal W}^{(t_2-t_1)} \vert \vec S(t_1) \rangle \frac{  P_{t_1}(\vec S(t_1)) }{P_{t_2}(\vec S(t_2)) }
=  \frac{ \langle \vec S(t_2) \vert  {\cal W}^{(t_2-t_1)} \vert \vec S(t_1) \rangle 
\langle \vec S(t_1) \vert {\cal W}^{t_1} \vert P^{ini} \rangle
 }{\langle \vec S(t_2) \vert {\cal W}^{t_2} \vert P^{ini} \rangle }
\label{WBackward propagator}
\end{eqnarray}
involves the three forward propagators $\langle \vec S(t_2) \vert  {\cal W}^{(t_2-t_1)} \vert \vec S(t_1) \rangle $, $\langle \vec S(t_1) \vert {\cal W}^{t_1} \vert P^{ini} \rangle $ and $ \langle \vec S(t_2) \vert {\cal W}^{t_2} \vert P^{ini} \rangle$.
The expression of Eq. \ref{WBackward propagator}
 can be considered as the direct generalization of the reweighting formula of Eq. \ref{WBchain} concerning a single time step with $t_1=t$ and $t_2=t+1$,
 while the backward propagator ${\cal B}_{0,T}[ \vec S(0) \vert \vec S(T) ]  $ for the special case $t_1=0$ and $t_2=T$ 
 reduces to
 \begin{eqnarray}
{\cal B}_{0,T}[ \vec S(0) \vert \vec S(T) ] 
=  \frac{ \langle \vec S(T) \vert  {\cal W}^{T} \vert \vec S(0) \rangle P^{ini} ( \vec S(0)) }
{\langle \vec S(T) \vert {\cal W}^{T} \vert P^{ini} \rangle }
 =  \frac{ \langle \vec S(T) \vert  {\cal W}^{T} \vert \vec S(0) \rangle P^{ini} ( \vec S(0)) }
{ \displaystyle \sum_{\vec \sigma} \langle \vec S(T) \vert  {\cal W}^{T} \vert \vec \sigma \rangle P^{ini} ( \vec \sigma )  }
\label{WBackward propagatorfull}
\end{eqnarray}
that only involves the forward propagator $\langle \vec S(T) \vert  {\cal W}^{T} \vert \vec \sigma \rangle $ over the time-window $T$
and the initial condition $P^{ini}(.) $.


\subsubsection{ Generated probability $B^{Gene}_t ( \vec S(t)) $ when the backwards dynamics is applied to $ P_*(\vec S(T))$
 instead of $P_T(\vec S(T) )$ }

The main idea of generative Markov models is to consider the trajectories 
that are generated by the time-dependent backward generator 
when, in the trajectories probabilities of Eq. \ref{Ptrajbackward}, 
the true forward solution $P_T(\vec S(T) ) $ at time $T$ 
is replaced by the steady state $P_*(\vec S(T)) $ of the forward dynamics 
 \begin{eqnarray}
&& {\cal P}^{GeneTraj} [ \vec S(0 \leq t \leq T)] 
 = P_*(\vec S(T) )  \prod_{t=0}^{T-1} \langle \vec S(t) \vert W^B_{t+1/2}\vert \vec S(t+1) \rangle
 = \frac{P_*(\vec S(T) )}{P_T(\vec S(T) )} {\cal P}^{Traj} [ \vec S(0 \leq t \leq T)] 
\label{PtrajbackwardGene}
\end{eqnarray}
So the probability ${\cal P}^{Traj} [ \vec S(0 \leq t \leq T)] $ of the forward trajectory $\vec S(0 \leq t \leq T) $ 
is reweighted by the ratio $ \frac{P_*(\vec S(T) )}{P_T(\vec S(T) )}$ to obtain the probability ${\cal P}^{GeneTraj} [ \vec S(0 \leq t \leq T)] $ of this trajectory in the backward-generative dynamics.
The hope is that the slight difference between $P_*(\vec S(T) ) $ and $P_T(\vec S(T) ) $ for large $T$
will produce probabilities ${\cal P}^{GeneTraj} [ \vec S(0 \leq t \leq T)] $ of generated-backward trajectories
that are close to the probabilities $ {\cal P}^{Traj} [ \vec S(0 \leq t \leq T)]$ of the forward trajectories,
and will lead in particular to generated
images $\vec S(0) $ at time $t=0$ with a probability distribution sufficiently close to the initial distribution $P^{ini}(\vec S(0)) $.

The generated distribution $B^{Gene}_t ( \vec S(t)) $ at time $t$ 
is obtained from the application of the backward 
propagator ${\cal B}_{t,T}[ \vec S(t) \vert \vec S(T) ] $ of Eq. \ref{WBackward propagator}
to $P_*(\vec S(T) ) $ 
 \begin{eqnarray}
B^{Gene}_t ( \vec S(t))
&& \equiv \sum_{\vec S(T) }{\cal B}_{t,T}[ \vec S(t) \vert \vec S(T) ]  P_*(\vec S(T) ) 
 = P_{t}(\vec S(t)) \sum_{\vec S(T) }  P_*(\vec S(T) )
 \frac{ \langle \vec S(T) \vert  {\cal W}^{(T-t)} \vert \vec S(t) \rangle  }{P_{T}(\vec S(T)) }
\nonumber \\
&& = \langle \vec S(t) \vert {\cal W}^t \vert P^{ini} \rangle \sum_{\vec S(T) }  P_*(\vec S(T) )
 \frac{ \langle \vec S(T) \vert  {\cal W}^{(T-t)} \vert \vec S(t) \rangle 
  }{ \langle \vec S(T) \vert {\cal W}^{T} \vert P^{ini} \rangle }  
\label{PtbackwardGene}
\end{eqnarray}
and can be compared 
with the forward solution $ P_{t}(\vec S(t))= \langle \vec S(t) \vert {\cal W}^t \vert P^{ini} \rangle $ 
that appears explicitly as prefactor before the sum over the configuration $\vec S(T) $.

For the special case $t=0$, the generated distribution of Eq. \ref{PtbackwardGene} 
 \begin{eqnarray}
B^{Gene}_0 ( \vec S(0))
&& =  \sum_{\vec S(T) }{\cal B}_{0,T}[ \vec S(0) \vert \vec S(T) ]  P_*(\vec S(T) ) 
= P^{ini}(\vec S(0)) \sum_{\vec S(T) }  P_*(\vec S(T) )
 \frac{ \langle \vec S(T) \vert  {\cal W}^T \vert \vec S(0) \rangle  }{P_{T}(\vec S(T)) }
\nonumber \\
&& = P^{ini}(\vec S(0)) \sum_{\vec S(T) }  P_*(\vec S(T) )
 \frac{ \langle \vec S(T) \vert  {\cal W}^T \vert \vec S(0) \rangle  }{\langle \vec S(T) \vert {\cal W}^{T} \vert P^{ini} \rangle }
\label{PtbackwardGenezero}
\end{eqnarray}
can be compared with the initial distribution $P^{ini}(\vec S(0)) $ 
that appears explicitly as prefactor before the sum over the configuration $\vec S(T) $.


\subsubsection{ Discussion }

In conclusion, all the properties of the non-trivial backward dynamics can be rewritten in terms of the initial condition $P^{ini}$ and in terms of the forward propagator, in particular:

(i)  at the level of a single time-step with the reweighting formula of Eq. \ref{WBchain}
for the time-dependent generator $\langle \vec S(t) \vert {\cal W}^B_{t+1/2}\vert \vec S(t+1) \rangle $;

(ii) at the level of any finite-time-interval $[t_1,t_2] \subset [0,T]$ with the backward propagator ${\cal B}_{t_1,t_2}[ \vec S(t_1) \vert \vec S(t_2) ]  $ of Eq. \ref{WBackward propagator};

(iii)  at the level of the probabilities of the whole trajectories $\vec S(0 \leq t \leq T) $ on the time-window $[0,T]$ with the reweighting formula of Eq. \ref{PtrajbackwardGene}.

For any time $t \in [0,T]$, the discrepancy between the generated distribution $B^{Gene}_t ( \vec S(t)) $
and the forward solution $ P_{t}(\vec S(t))= \langle \vec S(t) \vert {\cal W}^t \vert P^{ini} \rangle $ can be analyzed via Eq. \ref{PtbackwardGene} that simplifies into Eq. \ref{PtbackwardGenezero} for the special case $t=0$ where 
one can compare the probability $B^{Gene}_0 ( \vec S(0)) $ of the generated images
with the initial distribution $P^{ini}(\vec S(0)) $.

After this summary of the general framework, let us now 
focus on the explicit choice of the forward dynamics.


\subsection{ Explicit choice for the trivial forward noising dynamics}

\label{subsec_choice}

Since the forward dynamics should converge towards some simple steady state $P_*(\vec S)$ representing 'noise'
that can be easily generated,
 the simplest choice is when the $N$ pixels evolve towards 'noise' independently,
 i.e. when the forward propagator $ \langle \vec S(t) \vert {\cal W} \vert \vec S(t-1) \rangle $  is factorized 
 over the propagators $\langle S_n(t) \vert W \vert S_n(t-1) \rangle $ associated to the $N$ pixels $n=1,..,N$
 \begin{eqnarray}
 \langle \vec S(t) \vert {\cal W} \vert \vec S(t-1) \rangle && 
=  \prod_{n=1}^N \langle S_n(t) \vert W \vert S_n(t-1) \rangle 
\label{Wfactorized}
\end{eqnarray}
Note that in the field of stochastic dynamics of many-body spin or particle systems
where the main goal is to analyze the dynamical consequences of the interactions, 
this choice of a non-interacting dynamics is actually never considered and would appear as too trivial to be interesting on its own,
while from the point of view of generative Markov models, this choice of a trivial forward dynamics  
is fully justified by the further
purpose of constructing the non-trivial backward dynamics where the forward propagator plays the role of a simple building block,
while the non-trivial part enters via the initial distribution $P^{ini}$ of the interesting images.

\subsubsection{ Choice of the $2 \times 2$ Markov matrix $W$ governing the forward dynamics of a single pixel towards noise}

The simplest way to choose the $2 \times 2$ Markov matrix $W$ concerning a single pixel $S=\pm$
amounts to
use the well-known orthonormalized eigenvectors associated to the two eigenvalues $(\pm)$
of the Pauli matrix $\sigma^x$ 
  \begin{eqnarray}
 \vert \psi_0 \rangle  \equiv \vert \sigma^x=+1 \rangle && = \frac{\vert + \rangle +\vert - \rangle }{\sqrt 2}
\nonumber \\
\vert \psi_1 \rangle  \equiv \vert \sigma^x=-1 \rangle && = \frac{\vert + \rangle -\vert - \rangle }{\sqrt 2}
\label{paulisigmaxeigen}
\end{eqnarray}
in order to write the spectral decomposition of $W$ involving the largest eigenvalue unity $\lambda_0=1$ 
that is required by the probability conservation and 
the other smaller eigenvalue $\lambda \in ]0,1[$ that will parametrize the characteristic relaxation time towards pure noise
  \begin{eqnarray}
W  =   \vert \psi_0 \rangle \langle  \psi_0 \vert 
+ \lambda \vert \psi_1 \rangle \langle  \psi_1 \vert 
=\begin{pmatrix} 
\frac{1+\lambda}{2} &  \frac{1-\lambda}{2} \\
\frac{1-\lambda}{2} & \frac{1+\lambda}{2}
 \end{pmatrix}
  =  \frac{1+\lambda}{2} \mathbb{1} + \frac{1- \lambda}{2}  \sigma^x
\label{Wchainspectral}
\end{eqnarray}
The physical interpretation is 
that during one-time-step, the spin $S$ will flip towards $(-S)$ with the probability $\frac{1-\lambda}{2}$
and will keep its value $S$ with the complementary probability $\frac{1+\lambda}{2} $
  \begin{eqnarray}
  \langle + \vert W \vert - \rangle = \langle - \vert W \vert + \rangle&& = \frac{1-\lambda}{2} 
 \\
 \langle + \vert W \vert + \rangle = \langle - \vert W \vert - \rangle&& = \frac{1+\lambda}{2} 
\label{Wchainspectralelements}
\end{eqnarray}

The power $t$ of the Markov matrix $W$ of Eq. \ref{Wchainspectral} 
  \begin{eqnarray}
W^t 
 =\vert \psi_0 \rangle \langle \psi_0 \vert + \lambda^t \vert \psi_1 \rangle \langle \psi_1 \vert
 = \begin{pmatrix} 
\frac{ 1+\lambda^t  }{2} &  \frac{ 1-\lambda^t  }{2} \\
\frac{ 1-\lambda^t  }{2} & \frac{ 1+\lambda^t  }{2} 
 \end{pmatrix} =  \frac{1+\lambda^t}{2} \mathbb{1} + \frac{1- \lambda^t}{2}  \sigma^x
\label{Wchainspectralpower}
\end{eqnarray}
leads to the propagator $\langle S \vert W^t \vert S_0 \rangle $ between the value $S_0$ at time $t=0$
and the value $S$ at time $t$
 \begin{eqnarray}
\langle S \vert W^t \vert S_0 \rangle && = 
\frac{ 1+\lambda^t  }{2} \delta_{S,S_0}
+ \frac{ 1- \lambda^t }{2} \delta_{S,-S_0}
\label{Propagator1spin}
\end{eqnarray}
For any initial condition $S_0$, this propagator 
converges for large time $t \to + \infty$ 
towards the steady state $P_*(S=\pm)=\frac{1}{2}$ 
where the spin takes the two values $(\pm)$
with equal probabilities $(1/2,1/2)$
 \begin{eqnarray}
\langle S \vert W^t \vert S_0 \rangle \opsimeq_{t \to + \infty}   \frac{ \delta_{S,S_0} +\delta_{S,-S_0} }{2} = \frac{1}{2}\equiv P_*(S=\pm)
\label{Steady1spin}
\end{eqnarray}


\subsubsection{ Forward propagator $ \langle \vec S \vert {\cal W}^t \vert \vec \sigma \rangle $ for the $N$ pixels 
in terms of the overlap $Q$ between the two configurations $\vec S $ and $\vec \sigma $ } 

The forward propagator $ \langle \vec S \vert {\cal W}^t \vert \vec \sigma \rangle $ of Eq. \ref{Wfactorized} for the $N$ pixels
reduces to the product over the $N$ pixels of the single-spin-propagator of Eq. \ref{Propagator1spin}
 \begin{eqnarray}
 \langle \vec S \vert {\cal W}^t \vert \vec \sigma \rangle  
=  \prod_{n=1}^N \langle S_n \vert W^t \vert \sigma_n \rangle 
&& =  \prod_{n=1}^N  \left( \frac{ 1+\lambda^t  }{2} \delta_{S_n,\sigma_n}
+ \frac{ 1- \lambda^t }{2} \delta_{S_n,-\sigma_n}\right)
\nonumber \\
&& =    \left( \frac{ 1+\lambda^t  }{2} \right)^{ \displaystyle \sum_{n=1}^N \delta_{S_n,\sigma_n} }
\left( \frac{ 1- \lambda^t }{2} \right)^{\displaystyle \sum_{n=1}^N \delta_{S_n,-\sigma_n} }
\nonumber \\
&&= \left( \frac{ \sqrt{ 1-\lambda^{2t} } }{2} \right)^N
\left( \sqrt{\frac{1+\lambda^t }{1- \lambda^t} } \right)^{\displaystyle \sum_{n=1}^N S_n \sigma_n}
\label{Propagatorfactorizedimage}
\end{eqnarray}
The equivalence between the three alternative expressions is based on the fact that each spin can only take the two values $(\pm)$.
The two last expressions show the key role played by the extensive overlap $Q$ between the image $\vec S $ at time $t$ and the image $\vec \sigma $ at time $t=0$
  \begin{eqnarray}
 Q \equiv    \sum_{n=1}^N S_n \sigma_n = Q_+ - Q_-
\label{overlapQ}
\end{eqnarray}
with the two contributions $Q_{\pm}$ counting the numbers of pixels $n$ that have the same value or not 
in the two images
 \begin{eqnarray}
Q_+ \equiv \sum_{n=1}^N \delta_{S_n,\sigma_n}  = \frac{N+Q}{2}
\nonumber \\
Q_- \equiv \sum_{n=1}^N \delta_{S_n,-\sigma_n}  = \frac{N-Q}{2}
\label{Qpm}
\end{eqnarray}

Note that the overlap $Q$ of Eq. \ref{overlapQ} has the same parity as the number $N$ of pixels
and can take only the following $(N+1)$ values
\begin{eqnarray}
Q   \in \{-N,-N+2,-N+4,..,,...N-4,N-2,N\}
\label{Qeven}
\end{eqnarray}
In the whole paper, any sum over the overlap $Q$ means that the sum is over the possible values of Eq. \ref{Qeven}.

For large time $t \to + \infty$, the forward propagator of Eq. \ref{Propagatorfactorizedimage}
converges for any initial condition $\vec \sigma$ towards the steady distribution
 \begin{eqnarray}
 P_*( \vec S )= \frac{1}{2^N}
\label{steady}
\end{eqnarray}
where the possible $2^N$ configurations have all the same probability $ \frac{1}{2^N} $.


\subsection{ Direct link with the continuous-time Markov spin-flip dynamics of \cite{c_generativePropagator}}

\label{subsec_linkContoinuous}

The finite-time propagator of Eq. \ref{Propagatorfactorizedimage} 
obtained above for the discrete-time Markov chain dynamics 
has exactly the same form as the finite-time propagator written in Eq. 89 of \cite{c_generativePropagator}
for the continuous-time Markov spin-flip dynamics up to the correspondance of notations $\lambda = e^{-e_1 }$.
As a consequence, all the results that are based on this explicit finite-time propagator
can be directly translated between the discrete-time and the continuous-time frameworks.
While the work \cite{c_generativePropagator} focuses on the temporal convergence properties 
via the spectral decomposition of the forward propagator of Eq. \ref{Propagatorfactorizedimage}
with its consequences for observables like the local magnetizations 
${\cal M}_i(t) $ and the correlations ${\cal C}_{i_1,i_2,..,i_K}(t) $ between an arbitrary number $K$ of spins
$1 \leq i_1<i_2<..,i_K \leq N$
 \begin{eqnarray}
 {\cal M}_i(t) && = \sum_{\vec S} S_i \langle \vec S \vert {\cal W}^t \vert \vec \sigma \rangle  
 \nonumber \\
{\cal C}_{i_1,i_2,..,i_K}(t) && = \sum_{\vec S} \left( \prod_{k=1}^K S_{i_k}\right) \langle \vec S \vert {\cal W}^t \vert \vec \sigma \rangle  
\label{correforward}
\end{eqnarray}
the goal of the following section will be to analyze the full probability distributions of extensive global variables
corresponding to sums over the $N$ pixels: we have already stressed that the
overlap $Q$ of Eq. \ref{overlapQ} is the relevant variable that determines the forward propagator of Eq. \ref{Propagatorfactorizedimage}, but we will also consider the global magnetization 
  \begin{eqnarray}
 M \equiv    \sum_{n=1}^N S_n 
\label{MagneM}
\end{eqnarray}
that represents the simplest extensive variable associated to a single configuration.
When the number $N$ of pixels becomes large $N \to + \infty$, we will be interested into
 the large deviations properties of the corresponding intensive variables $q=\frac{Q}{N}$ and/or $m=\frac{M}{N}$
 during the forward dynamics and into the consequences for the backward-generative dynamics.


\section{ Large deviations properties of the forward propagator for $N$ pixels   }

\label{sec_jointMQ}

In this section, the goal is to analyze the statistical properties of the forward propagator $ \langle \vec S \vert {\cal W}^t \vert \vec \sigma \rangle $ of Eq. \ref{Propagatorfactorizedimage} for any finite number $N$ of pixels
and to extract the large deviations properties in the limit $N \to + \infty$.


\subsection{ Distribution $P_t(Q)$ of the overlap $Q$ with the initial condition }

 The probability distribution $P_t(Q) $ of the overlap $Q$ at time $t$ produced by the forward propagator $ \langle \vec S \vert {\cal W}^t \vert \vec \sigma \rangle $ of Eq. \ref{Propagatorfactorizedimage} 
 \begin{eqnarray}
P_t (Q) && \equiv \sum_{\vec S}   \delta_{Q,  \sum_{n=1}^N S_n \sigma_n}\langle \vec S \vert {\cal W}^t \vert \vec \sigma \rangle  
 =\left( \frac{ 1+\lambda^t  }{2} \right)^{ \frac{N+Q}{2} }
\left( \frac{ 1- \lambda^t }{2} \right)^{\frac{N-Q}{2} }
  \sum_{\vec S}  \delta_{Q,  \sum_{n=1}^N S_n \sigma_n}
\nonumber \\
&& =\left( \frac{ 1+\lambda^t  }{2} \right)^{ \frac{N+Q}{2} }
\left( \frac{ 1- \lambda^t }{2} \right)^{\frac{N-Q}{2} }
  \frac{ N!}{\frac{N +Q}{2}  !  \frac{N -Q}{2} !}
\label{ProbaQtalone}
\end{eqnarray}
 reduces to the binomial distribution.

For large $N$, the Stirling approximation for the logarithm of the factorial 
 \begin{eqnarray}
\ln \left[ (N x)! \right] \opsimeq_{N \to + \infty} 
x N   \ln N 
+ N x  \ln x 
- N x  
+  \frac{1}{2} \ln (2 \pi N)
+ \frac{1}{2} \ln  x
+ O\left( \frac{1}{N} \right)
\label{stirling}
\end{eqnarray}
 leads to the following large deviation behavior the intensive overlap $q=\frac{Q}{N}$
 \begin{eqnarray}
P_t (Q=N q) \oppropto_{N \to + \infty} e^{- N i_t (q)  } 
\label{ProbaQtalonelargedev}
\end{eqnarray}
with the standard rate function $i_t(q)$ produced by the binomial distribution
 \begin{eqnarray}
i_t (q) =
 \frac{ 1+q }{2}  \ln \left( \frac{1+q  }{ 1+\lambda^t }\right)  
+  \frac{ 1-q }{2}  \ln \left( \frac{1-q  }{ 1-\lambda^t }\right)   
\label{rateitqdirect}
\end{eqnarray}

At the typical value ${\hat q}=\lambda^t$, the rate function $i_t (q) $ vanishes $i_t ({\hat q}) =0 $
together with its first derivative 
 \begin{eqnarray}
\frac{ d i_t (q) }{dq} =  \frac{ 1 }{2}  \ln \left( \frac{1+q  }{ 1+\lambda^t }\right)  
-  \frac{ 1 }{2}  \ln \left( \frac{1-q  }{ 1-\lambda^t }\right)  
\label{rateitqdirectderi}
\end{eqnarray}

In the following subsection, the goal is to generalize this simple analysis to take into account the magnetization.


\subsection{ Joint distribution $P_t^{[\vec \sigma]}(M,Q)$ of the magnetization $M$ and of the overlap $Q$ 
with the initial condition $\vec \sigma$}

\subsubsection{ Explicit joint distribution $P_t^{[\vec \sigma]}(M,Q) $ for any finite number $N$ of pixels }

For a given initial condition $\vec \sigma$, the joint distribution $P_t^{[\vec \sigma]}(M,Q) $ of the magnetization $M$
and of the overlap $Q$ produced by the forward propagator $\vec S \vert {\cal W}^t \vert \vec \sigma \rangle $ at time $t$ of Eq. \ref{Propagatorfactorizedimage} reads
 \begin{eqnarray}
P^{[\vec\sigma]}_t (M,Q) && \equiv \sum_{\vec S} \delta_{M, \sum_{n=1}^N S_n}  \delta_{Q,  \sum_{n=1}^N S_n \sigma_n}\langle \vec S \vert {\cal W}^t \vert \vec \sigma \rangle  
\nonumber \\
&& =  \left( \frac{ 1+\lambda^t  }{2} \right)^{ \frac{N+Q}{2} }
\left( \frac{ 1- \lambda^t }{2} \right)^{\frac{N-Q}{2} }
\sum_{\vec S} \delta_{M, \sum_{n=1}^N S_n}  \delta_{Q,  \sum_{n=1}^N S_n \sigma_n}
\label{JointMQ}
\end{eqnarray}

Since the initial condition $\vec \sigma$ is fixed,
its magnetization $M_0$ 
\begin{eqnarray}
M_0 \equiv \sum_{n=1}^N\sigma_n
\label{M0initial}
\end{eqnarray}
determines the corresponding populations of up and down spins in the initial condition
 \begin{eqnarray}
N_+ \equiv  \sum_{n=1}^N  \delta_{\sigma_n,1} = \frac{N+M_0}{2}
\nonumber \\
N_- \equiv  \sum_{n=1}^N  \delta_{\sigma_n,-1} = \frac{N-M_0}{2}
  \label{2popM0}
\end{eqnarray}

In order to analyze the joint properties of $M$ and $Q$, one needs
to introduce the four populations
labelled by $\epsilon=\pm$ and $\epsilon_0=\pm$
\begin{eqnarray}
 N^{\epsilon}_{\epsilon_0} && =\sum_{n=1}^N  \delta_{S_n,\epsilon} \delta_{\sigma_n,\epsilon_0}
 =  \frac{ N+ \epsilon_0 M_0 +\epsilon M   +\epsilon \epsilon_0 Q}{4}
\label{4populations}
\end{eqnarray}
with the sum rules
\begin{eqnarray}
\sum_{\epsilon=\pm} \sum_{\epsilon_0=\pm}   N^{\epsilon}_{\epsilon_0} && =  N
 \nonumber \\
\sum_{\epsilon=\pm} \sum_{\epsilon_0=\pm} \epsilon  N^{\epsilon}_{\epsilon_0} && =  M
 \nonumber \\
 \sum_{\epsilon=\pm} \sum_{\epsilon_0=\pm} \epsilon_0   N^{\epsilon}_{\epsilon_0} && =  M_0
 \nonumber \\
\sum_{\epsilon=\pm} \sum_{\epsilon_0=\pm} \epsilon \epsilon_0   N^{\epsilon}_{\epsilon_0} && =  Q
\label{sumrules}
\end{eqnarray}

Then the fixed population $N_+$ of Eq. \ref{2popM0} can be decomposed into the two subpopulations
\begin{eqnarray}
N^+_+ &&  \equiv  \sum_{n=1}^N  \delta_{S_n,1} \delta_{\sigma_n,1} =  \frac{ N +  M_0 + M +  Q}{4}
\nonumber \\
N^-_+  && \equiv  \sum_{n=1}^N  \delta_{S_n,-1} \delta_{\sigma_n,1} = \frac{ N+ M_0 - M   -  Q}{4}
  \label{2subpop+}
\end{eqnarray}
while the fixed population $N_-$ of Eq. \ref{2popM0} can be decomposed into the two subpopulations
\begin{eqnarray}
N^+_-  && \equiv  \sum_{n=1}^N  \delta_{S_n,1} \delta_{\sigma_n,-1} = \frac{ N- M_0 + M   -  Q}{4}
\nonumber \\
N^-_- &&  \equiv  \sum_{n=1}^N  \delta_{S_n,-1} \delta_{\sigma_n,-1} = \frac{ N- M_0 - M   +  Q}{4}
  \label{2subpop-}
\end{eqnarray}

As a consequence, the joint distribution of Eq. \ref{JointMQ} 
depends on the initial condition $\vec \sigma$ only via its magnetization $M_0$
and involves the two binomial coefficients associated to the combinatorics of the two decompositions of Eqs \ref{2subpop+}
and \ref{2subpop-}
 \begin{eqnarray}
P^{[M_0]}_t (M,Q) && =  
\left( \frac{ 1+\lambda^t  }{2} \right)^{ \frac{N+Q}{2} }
\left( \frac{ 1- \lambda^t }{2} \right)^{\frac{N-Q}{2} } \times
 \frac{ N_+ !}{N^+_+! N^-_+!}   \times  \frac{ N_- !}{N^+_-! N^-_-!}
\nonumber \\
&&  = \left( \frac{ 1+\lambda^t  }{2} \right)^{ \frac{N+Q}{2} }
\left( \frac{ 1- \lambda^t }{2} \right)^{\frac{N-Q}{2} }
\frac{ \frac{N +M_0}{2}  !  \frac{N -M_0}{2} !} 
 {\frac{N +  M_0 + M +  Q}{4}  !  \frac{ N+ M_0 - M   -  Q }{4}!\frac{N- M_0 + M   -  Q}{4}!  \frac{ N- M_0 - M   +  Q }{4}!}
\label{JointMQbino}
\end{eqnarray}


\subsubsection{ Large deviations for large $N$ : rate function ${\cal I}_t^{[m_0]} (m,q) $ for the intensive magnetization $m$ and the intensive overlap $q$}

For large $N$, one is interested into the behavior of the joint distribution of Eq. \ref{JointMQbino}
 \begin{eqnarray}
P^{[M_0=N m_0]}_t (M=N m,Q=N q) && 
 = \left( \frac{ 1+\lambda^t  }{2} \right)^{ N \frac{1+q}{2} }
\left( \frac{ 1- \lambda^t }{2} \right)^{N \frac{1-q}{2} }
 \frac{ \left( N \frac{1+m_0}{2} \right) !  \left( N \frac{1-m_0}{2} \right) ! } 
 { \displaystyle \prod_{\epsilon=\pm} \prod_{\epsilon_0=\pm} \left( N \frac{1+\epsilon m + \epsilon_0 m_0+\epsilon \epsilon_0 q}{4} \right)  ! }
\label{JointMQbinorescal}
\end{eqnarray}
as a function of the two intensive magnetizations $m$ and $m_0$ and of the intensive overlap $q$
\begin{eqnarray}
m_0 \equiv \frac{M_0}{N} = \frac{1}{N}\sum_{n=1}^N\sigma_n
\nonumber \\
m \equiv \frac{M}{N} = \frac{1}{N}\sum_{n=1}^NS_n
\nonumber \\
q \equiv \frac{Q}{N} = \frac{1}{N}\sum_{n=1}^NS_n \sigma_n
\label{intensive}
\end{eqnarray}

As described in more details in the Appendix section \ref{app_stirling},
Eq. \ref{JointMQbinorescal}
displays the following large deviation behavior for large $N$
 \begin{eqnarray}
P^{[M_0=N m_0]}_t (M=N m,Q=N q) \oppropto_{N \to + \infty} e^{- N  {\cal I}_t^{[m_0]}( m,q)  } 
\label{JointMQbinorescallargedev}
\end{eqnarray}
with the rate function 
 \begin{eqnarray}
  {\cal I}_t^{[m_0]}( m,q)  && = 
 - \frac{1+q}{2} \ln \left( \frac{ 1+\lambda^t  }{2} \right)
- \frac{1-q}{2}  \ln \left( \frac{ 1- \lambda^t }{2} \right)
- \sum_{\eta=\pm 1}\frac{1+\eta m_0}{2}  \ln \left( \frac{1+\eta m_0}{2} \right)   
\nonumber \\
&& +   \sum_{\epsilon=\pm 1}\sum_{\epsilon_0=\pm 1}  \frac{1+\epsilon m + \epsilon_0 m_0+\epsilon \epsilon_0 q}{4}  \ln \left( \frac{1+\epsilon m + \epsilon_0 m_0+\epsilon \epsilon_0 q}{4}\right)
\label{rateImq}
\end{eqnarray}

Let us compute the first-order partial derivatives 
with respect to the magnetization $m$
 \begin{eqnarray}
\partial_m  {\cal I}_t^{[m_0]}( m,q)  
&& =   \sum_{\epsilon=\pm 1}\sum_{\epsilon_0=\pm 1} 
 \frac{\epsilon }{4}  \ln \left( 1+\epsilon m + \epsilon_0 m_0+\epsilon \epsilon_0 q\right)
= \frac{1 }{4}  \ln \left( \frac{(1+ m)^2 - (  m_0+  q)^2}{(1-m)^2 -(m_0-q)^2}\right)
\label{rateImqderim}
\end{eqnarray}
and with respect to the overlap $q$
 \begin{eqnarray}
 \partial_q  {\cal I}_t^{[m_0]}( m,q) 
 &&  = 
 - \frac{1}{2} \ln \left( \frac{ 1+\lambda^t  }{2} \right)
+ \frac{1}{2}  \ln \left( \frac{ 1- \lambda^t }{2} \right)
 +   \sum_{\epsilon=\pm 1}\sum_{\epsilon_0=\pm 1}  \frac{\epsilon \epsilon_0 }{4} 
  \ln \left( 1+\epsilon m + \epsilon_0 m_0+\epsilon \epsilon_0 q\right)
  \nonumber \\
&& =  - \frac{1}{2} \ln \left( \frac{ 1+\lambda^t  }{1-\lambda^t } \right)
+ \frac{ 1 }{4}  \ln \left( \frac{(1+q)^2- (m +  m_0)^2}{(1-q)^2 - (m - m_0)^2}\right) 
\label{rateImqderiq}
\end{eqnarray}
as well as the second-order partial derivatives 
 \begin{eqnarray}
\partial_m^2  {\cal I}_t^{[m_0]}( m,q)  = \partial_q^2  {\cal I}_t^{[m_0]}( m,q) 
&& = \frac{1}{4}  \sum_{\epsilon=\pm 1}\sum_{\epsilon_0=\pm 1}  \frac{1 }{1+\epsilon m + \epsilon_0 m_0+\epsilon \epsilon_0 q}
 \nonumber \\ 
 && = \frac{1}{2} \left[ \frac{ 1+m_0}{ (1+m_0)^2 - (m+q)^2 } + \frac{ 1-m_0}{ (1-m_0)^2 - (m-q)^2 }\right]
\label{rateImqderimm}
\end{eqnarray}
and
 \begin{eqnarray}
\partial_q \partial_m  {\cal I}_t^{[m_0]}( m,q)  
&& =  \frac{1}{4}  \sum_{\epsilon=\pm 1}\sum_{\epsilon_0=\pm 1}  \frac{\epsilon_0 }{1+\epsilon m + \epsilon_0 m_0+\epsilon \epsilon_0 q} 
 \nonumber \\ 
 && = \frac{1}{2} \left[ \frac{ 1+m_0}{ (1+m_0)^2 - (m+q)^2 } - \frac{ 1-m_0}{ (1-m_0)^2 - (m-q)^2 }\right]
\label{rateImqderimqcross}
\end{eqnarray}

At the typical value $(\hat m,\hat q)$ of the pair $(m,q)$, 
the vanishing of the first derivatives of Eqs \ref{rateImqderim}
and \ref{rateImqderiq}
 \begin{eqnarray}
\partial_m  {\cal I}_t^{[m_0]}( m,q)  =0 =\partial_q  {\cal I}_t^{[m_0]}( m,q) \text{ for } \ ( m,q)=(\hat m,\hat q)
\label{bothderivanish}
\end{eqnarray}
leads to the expected solution 
 \begin{eqnarray}
  \hat q && = \lambda^t
\nonumber \\
\hat m && = \lambda^t m_0
\label{bothderivanisheq}
\end{eqnarray}
where the rate function ${\cal I}_t^{[m_0]}( m,q) $ itself also vanishes 
 \begin{eqnarray}
{\cal I}_t^{[m_0]}( \hat m =\lambda^t m_0, \hat q=\lambda^t) =0
\label{Imqtypvanish}
\end{eqnarray}
 
The second derivatives of Eqs \ref{rateImqderimm} and \ref{rateImqderimqcross} evaluated for $(m,q)=(\hat m,\hat q)$
 \begin{eqnarray}
\partial_m^2  {\cal I}_t^{[m_0]}( m,q) \big\vert_{(m,q)=(\hat m,\hat q)} 
= \partial_q^2  {\cal I}_t^{[m_0]}( m,q) \big\vert_{(m,q)=(\hat m,\hat q)}
  && = \frac{1}{ (1-m_0^2) (1-\lambda^{2t} )}
\nonumber \\
\partial_q \partial_m  {\cal I}_t^{[m_0]}( m,q)  \big\vert_{(m,q)=(\hat m,\hat q)} 
 && = - \frac{  m_0 }{ (1-m_0^2) (1-\lambda^{2t} )}
\label{simplideriforgauss}
\end{eqnarray}
lead to the following Taylor series of the rate function $ {\cal I}_t^{[m_0]}( m,q)$
at second order around the typical value $( \hat m =\lambda^t m_0, \hat q=\lambda^t)$ of the pair $(m,q)$
where the function and the two first derivatives vanish (Eqs \ref{bothderivanish} and \ref{Imqtypvanish})
 \begin{eqnarray}
  {\cal I}_t^{[m_0]}( m,q)   
 =  \frac{(m-  \lambda^t m_0)^2+(q-\lambda^t )^2 - 2 m_0 (m-  \lambda^t m_0)(q-\lambda^t )}{2 (1-m_0^2) (1-\lambda^{2t} )} +...
\label{taylorgauss}
\end{eqnarray}
that governs the joint Gaussian typical fluctuations of order $\frac{1}{\sqrt{N} }$ 
for the pair $(m,q)$ around the typical value $( \hat m =\lambda^t m_0, \hat q=\lambda^t)$
 \begin{eqnarray}
 p^{[m_0]}_t( m,q)    \oppropto_{N \to + \infty} e^{\displaystyle  - N {\cal I}_t^{[m_0]}( m,q)}
 \oppropto_{N \to + \infty}   e^{\displaystyle  - N \frac{(m-  \lambda^t m_0)^2+(q-\lambda^t )^2 - 2 m_0 (m-  \lambda^t m_0)(q-\lambda^t )}{2 (1-m_0^2) (1-\lambda^{2t} )} +...}
\label{taylorgaussapproxintermediate}
\end{eqnarray}
The prefactor of this bivariate Gaussian distribution can be reconstructed from the normalization condition 
when one integrates over the two variables $(m,q)$
in order to obtain the following Gaussian approximation for large $N$
 \begin{eqnarray}
 p^{[m_0]Gauss}_t( m,q)   
 = \frac{1}{2 \pi (1-\lambda^{2t}) (1-m_0^2) }  e^{\displaystyle  - N \frac{(m-  \lambda^t m_0)^2+(q-\lambda^t )^2 - 2 m_0 (m-  \lambda^t m_0)(q-\lambda^t )}{2 (1-m_0^2) (1-\lambda^{2t} )} }
\label{taylorgaussapprox}
\end{eqnarray}



\subsection{ Generating function ${\cal G}^{[\vec\sigma]}_t (h,g) $ of the magnetization $M$ and of the overlap $Q$ with the initial condition $\vec \sigma$ }

\subsubsection{ Generating function ${\cal G}^{[\vec\sigma]}_t (h,g) $ for any finite $N$ }

Instead of the joint probability distribution $P^{[\vec\sigma]}_t (M,Q) $ of Eq. \ref{JointMQ},
one can alternatively focus on the corresponding generating function ${\cal G}^{[\vec\sigma]}_t (h,g) $ involving the conjugated parameters $(h,g)$
 \begin{eqnarray}
{\cal G}^{[\vec\sigma]}_t (h,g) && \equiv \sum_M \sum_Q e^{hM+gQ } P^{[\vec\sigma]}_t (M,Q) 
\nonumber \\
&&
 = \sum_{\vec S} e^{ \displaystyle h \sum_{n=1}^N S_n  +g  \sum_{n=1}^N S_n \sigma_n}
\left( \frac{ \sqrt{ 1-\lambda^{2t} } }{2} \right)^N
\prod_{n=1}^N \left( \sqrt{\frac{1+\lambda^t }{1- \lambda^t} } \right)^{S_n \sigma_n}
\label{Jointgeneratingdef}
\end{eqnarray}

It is then convenient to introduce the notation
 \begin{eqnarray}
\mu_t && \equiv \frac{1}{2} \ln \left( \frac{1+\lambda^t }{1- \lambda^t}  \right) = \arctanh \lambda^t 
\nonumber \\
\lambda^t && = \tanh \mu_t
\label{mut}
\end{eqnarray}
with
 \begin{eqnarray}
\sqrt{ 1-\lambda^{2t}} =  \sqrt{1- \tanh^2 \mu_t} = \frac{ 1 }{ \cosh \mu_t }
\label{coshmut}
\end{eqnarray}
in order to evaluate Eq. \ref{Jointgeneratingdef}
 \begin{eqnarray}
{\cal G}^{[\vec\sigma]}_t (h,g) && = 
\left( \frac{ 1 }{2 \cosh \mu_t }\right)^N
\sum_{\vec S} e^{ \displaystyle h \sum_{n=1}^N S_n  +(g + \mu_t) \sum_{n=1}^N S_n \sigma_n} 
 = \left( \frac{ 1 }{2 \cosh \mu_t }\right)^N
\prod_{n=1}^N \left( \sum_{S_n=\pm 1} e^{ S_n \left[ h + (g+\mu_t ) \sigma_n \right]} \right)
\nonumber \\
&& = \left( \frac{ 1 }{2 \cosh \mu_t }\right)^N
\prod_{n=1}^N \left(  e^{ h + (g+\mu_t ) \sigma_n} +e^{ - h - (g+\mu_t ) \sigma_n} \right)
=\left( \frac{ 1 }{ \cosh \mu_t }  \right)^N
\prod_{n=1}^N  \cosh \left[ h + (g+\mu_t)  \sigma_n \right] 
\label{Jointgenerating}
\end{eqnarray}

Since the spin $\sigma_n$ takes only the two values $\sigma_n=\pm 1$,
one can rewrite
\begin{eqnarray}
  \cosh \left( h + (g+\mu_t)  \sigma_n \right) =
  \sqrt{ \cosh \left[ h + (g+\mu_t)   \right]  \cosh \left[ h - (g+\mu_t)   \right]}  
  \left( \sqrt{\frac{ \cosh \left[ h + (g+\mu_t)   \right] }{ \cosh \left[ h - (g+\mu_t)   \right]} } \right)^{ \sigma_n}
\label{coshexpand}
\end{eqnarray}
to obtain that Eq. \ref{Jointgenerating}
 \begin{eqnarray}
{\cal G}^{[\vec\sigma]}_t (h,g)  
&& =\left(  \frac{ \sqrt{ \cosh \left[ h + (g+\mu_t)   \right]  \cosh \left[ h - (g+\mu_t)   \right]} }{ \cosh \mu_t }  \right)^N
 \left( \sqrt{\frac{ \cosh \left[ h + (g+\mu_t)   \right] }{ \cosh \left[ h - (g+\mu_t)   \right]} } \right)^{ 
 \displaystyle  \sum_{n=1}^N\sigma_n}
 \nonumber \\
 && \equiv {\cal G}_t^{[M_0=\sum_{n=1}^N\sigma_n ]} (h,g)
\label{JointgeneratingM0}
\end{eqnarray}
depends on the initial condition $\vec \sigma$ only via its magnetization $M_0$ of Eq. \ref{M0initial},
in consistency with the same statement for the probability distribution of Eq. \ref{JointMQbino}.

The series expansion of ${\cal G}^{[\vec\sigma]}_t (h,g) $ of Eq. \ref{Jointgeneratingdef} 
with respect to the two variables $(h,g)$  
 \begin{eqnarray}
{\cal G}^{[M_0]}_t (h,g) && \equiv \sum_M \sum_Q e^{hM+gQ } P^{[\vec\sigma]}_t (M,Q) 
\nonumber \\
&&
 = 1 + \sum_M \sum_Q \left[  hM+gQ + \frac{h^2 M^2 + 2 hg MQ +g^2Q^2}{2} + ... \right] P^{[\vec\sigma]}_t (M,Q) 
\nonumber \\
&& \equiv 1+ h {\cal M}_1(t) + g {\cal Q}_1(t) + \frac{h^2}{2} {\cal M}_2(t)+ \frac{g^2}{2}{\cal Q}_2(t)
 + hg {\cal C}_{1,1}(t) +...
\label{Jointgeneratingdefmoments}
\end{eqnarray}
involves
the various moments and correlations of the joint distribution $P^{[\vec\sigma]}_t (M,Q) $
 \begin{eqnarray}
{\cal M}_k(t) && \equiv \sum_M \sum_Q M^k P^{[\vec\sigma]}_t (M,Q) 
\nonumber \\
{\cal Q}_l(t) && \equiv \sum_M \sum_Q Q^l P^{[\vec\sigma]}_t (M,Q) 
\nonumber \\
{\cal C}_{k,l}(t) && \equiv \sum_M \sum_Q M^k Q^l P^{[\vec\sigma]}_t (M,Q) 
\label{defmomentsMQ}
\end{eqnarray}
while the generating function of the various cumulants and connected correlations 
appear in the series expansion of its logarithm
 \begin{eqnarray}
 \ln \left( {\cal G}^{[\vec\sigma]}_t (h,g) \right) 
&& = \ln \left( 1+ h {\cal M}_1(t) + g {\cal Q}_1(t) + \frac{h^2}{2} {\cal M}_2(t)+ \frac{g^2}{2}{\cal Q}_2(t)
 + hg {\cal C}_{1,1}(t) +... \right)
\nonumber \\
&& = h {\cal M}_1(t) + g {\cal Q}_1(t) + \frac{h^2}{2} \left( {\cal M}_2(t)- {\cal M}_1^2(t)\right)
+ \frac{g^2}{2} \left( {\cal Q}_2(t) -{\cal Q}_1^2(t) \right)
 + hg \left( {\cal C}_{1,1}(t) - {\cal M}_1 {\cal Q}_1\right) 
  + ...
\label{logJointgeneratingdefcumul}
\end{eqnarray}


\subsubsection{ Large deviations for large $N$ : generating function ${\cal F} _t^{[m_0]} (h,g) $ of scaled cumulants }

For large $N$, the logarithm of the generating function $ {\cal G}_t^{[M_0=N m_0]} (h,g)$ of Eq. \ref{JointgeneratingM0}
is dominated by the extensive behavior with respect to the size $N$
 \begin{eqnarray}
\ln \left( {\cal G}^{[M_0=N m_0]}_t (h,g) \right)  \oppropto_{N \to + \infty}  N {\cal F} _t^{[m_0]} (h,g)
\label{JointgeneratingM0largedev}
\end{eqnarray}
where the generating function ${\cal F} _t^{[m_0]} (h,g) $ of the scaled cumulants 
 \begin{eqnarray}
{\cal F} _t^{[m_0]} (h,g) && = \lim_{N \to + \infty} \left[ \frac{ \ln \left( {\cal G}^{[\vec\sigma]}_t (h,g) \right) }{N} \right]
\label{logJointgeneratingdefcumulrescal}
 \\
&& =\lim_{N \to + \infty} \left[ h \frac{{\cal M}_1(t)}{N} + g \frac{{\cal Q}_1(t) }{N}
+ \frac{h^2}{2} \left( \frac{{\cal M}_2(t)- {\cal M}_1^2(t)}{N}\right)
+ \frac{g^2}{2} \left(\frac{ {\cal Q}_2(t) -{\cal Q}_1^2(t)}{N} \right)
 + hg \left( \frac{{\cal C}_{1,1}(t) - {\cal M}_1 {\cal Q}_1}{N}\right) 
  + ...
  \right]
\nonumber
\end{eqnarray}
reads
in terms of the intensive magnetization $m_0=\frac{M_0}{N}$ 
 \begin{eqnarray}
 {\cal F} _t^{[m_0]} (h,g) 
&&  = \ln \left(  \frac{ \sqrt{ \cosh \left[ h + (g+\mu_t)   \right]  \cosh \left[ h - (g+\mu_t)   \right]} }{ \cosh \mu_t }  \right)
 + m_0  \ln \left( \sqrt{\frac{ \cosh \left[ h + (g+\mu_t)   \right] }{ \cosh \left[ h - (g+\mu_t)   \right]} } \right)
 \nonumber \\
  && = - \ln (\cosh \mu_t)
  + \frac{1+m_0}{2}  \ln \left(  \cosh \left[ h + (g+\mu_t)   \right]  \right)
  + \frac{1-m_0}{2} \ln \left(  \cosh \left[ h - (g+\mu_t)   \right]  \right)
\label{scgf}
\end{eqnarray}

The first-order partial derivatives with respect to $h$ and with respect to $g$ are given by
\begin{eqnarray}
\partial_h {\cal F} _t^{[m_0]} (h,g) 
 &&  = \frac{1+m_0}{2}    \tanh \left[ h + (g+\mu_t)   \right] +\frac{1-m_0}{2} \tanh \left[ h - (g+\mu_t)   \right] 
\label{scgfderih}
\end{eqnarray}
and
\begin{eqnarray}
\partial_g {\cal F} _t^{[m_0]} (h,g) 
 &&  = \frac{1+m_0}{2}   \tanh \left[ h + (g+\mu_t)   \right] - \frac{1-m_0}{2}\tanh \left[ h - (g+\mu_t)   \right]
\label{scgfderig}
\end{eqnarray}
while the second-order partial derivatives reduce to
\begin{eqnarray}
\partial_h^2 {\cal F} _t^{[m_0]} (h,g) 
   = \frac{1}{    \cosh^2 \left[ h + (g+\mu_t)   \right] } = \partial_g^2 {\cal F} _t^{[m_0]} (h,g) 
\label{scgfderihh}
\end{eqnarray}
and
\begin{eqnarray}
\partial_g \partial_h {\cal F} _t^{[m_0]} (h,g) 
 &&  =  \frac{m_0}{    \cosh^2 \left[ h + (g+\mu_t)   \right] }
\label{scgfderihg}
\end{eqnarray}

In particular for $h=0=g$, the first-order derivatives correspond to the averaged values of the intensive magnetization
and of the intensive overlap that appear at first-order 
in the series expansion of Eq. \ref{logJointgeneratingdefcumulrescal}
\begin{eqnarray}
\partial_h {\cal F} _t^{[m_0]} (h,g) \bigg\vert_{h=0=g }
 &&  = m_0  \tanh \mu_t  = m_0 \lambda^t = \lim_{N \to + \infty} \left(  \frac{{\cal M}_1(t)}{N} \right) =\hat m
\nonumber \\
\partial_g {\cal F} _t^{[m_0]} (h,g) \bigg\vert_{h=0=g }
 &&  = \tanh \mu_t  = \lambda^t =\lim_{N \to + \infty} \left(  \frac{{\cal Q}_1(t)}{N} \right) =\hat q
\label{dericumulants}
\end{eqnarray}
that are in agreement with the typical values $( \hat m =\lambda^t m_0, \hat q=\lambda^t)$ of Eq. \ref{bothderivanisheq} as it should,
while the second-order derivatives 
lead to the rescaled variances of the magnetization and of the overlap, and to their rescaled connected correlation
 that appear at second-order 
in the series expansion of Eq. \ref{logJointgeneratingdefcumulrescal}
\begin{eqnarray}
\partial_h^2 {\cal F} _t^{[m_0]} (h,g) \bigg\vert_{h=0=g }
  && = \frac{1}{    \cosh^2 \left[ \mu_t   \right] } = 1-\lambda^{2t} 
  =   \lim_{N \to + \infty}  \left( \frac{{\cal M}_2(t)- {\cal M}_1^2(t)}{N}\right) 
  \nonumber \\
 \partial_g^2 {\cal F} _t^{[m_0]} (h,g) \bigg\vert_{h=0=g }
  && = \frac{1}{    \cosh^2 \left[ \mu_t   \right] } = 1-\lambda^{2t} 
  =     \lim_{N \to + \infty}  \left( \frac{{\cal Q}_2(t)- {\cal Q}_1^2(t)}{N}\right) 
  \nonumber \\
\partial_g \partial_h {\cal F} _t^{[m_0]} (h,g) \bigg\vert_{h=0=g }
 &&  =  \frac{m_0}{    \cosh^2 \left[ \mu_t   \right] } =  m_0 \left( 1-\lambda^{2t} \right) 
 =\lim_{N \to + \infty}  \left( \frac{{\cal C}_{1,1}(t) - {\cal M}_1 {\cal Q}_1}{N}\right) 
\label{scgfderihg0}
\end{eqnarray}
These values are in agreement with their alternative evaluations from the joint Gaussian approximation of Eq. \ref{taylorgaussapprox}
for the pair $(m,q)$ around the typical value $( \hat m =\lambda^t m_0, \hat q=\lambda^t)$.


\subsubsection{ Legendre transformation between the rate function ${\cal I}_t^{[m_0]}( m,q)  $ 
and the scaled cumulant generating function ${\cal F} _t^{[m_0]} (h,g) $ }

A central result in the theory of large deviations
is that the rate function and the scaled cumulant generating function 
are directly related via Legendre transformations, as we now recall for the present setting involving two variables. 

Plugging the large deviations form of Eq. \ref{JointMQbinorescallargedev}
for $P^{[M_0=N m_0]}_t (M=N m,Q=N q) $ 
into the generating function of Eq. \ref{Jointgeneratingdef} that displays the asymptotic behavior of Eq. \ref{JointgeneratingM0largedev}
 \begin{eqnarray}
{\cal G}^{[\vec\sigma]}_t (h,g) && \oppropto_{N \to + \infty}  
 \int dm \int dq e^{ N \left[ hm+gq - {\cal I}_t^{[m_0]}( m,q)   \right] }  
 \oppropto_{N \to + \infty}e^{ N {\cal F} _t^{[m_0]} (h,g)}
\label{Jointgeneratingsaddle}
\end{eqnarray}
yields, via the saddle-point evaluation for large $N \to + \infty$ of the integrals over the two variables $m$ and $q$,
that the scaled cumulant generating function ${\cal F} _t^{[m_0]} (h,g) $
is related to the rate function ${\cal I}_t^{[m_0]}( m,q)  $
via the double Legendre transform
that involves the first derivatives of Eqs \ref{rateImqderim} and \ref{rateImqderiq}
 \begin{eqnarray}
{\cal F} _t^{[m_0]} (h,g) && = hm+gq-  {\cal I}_t^{[m_0]}( m,q)  
\nonumber \\
h && =  \partial_m {\cal I}_t^{[m_0]}( m,q)  = \frac{1 }{4}  \ln \left( \frac{(1+ m)^2 - (  m_0+  q)^2}{(1-m)^2 -(m_0-q)^2}\right)
\nonumber \\
g && =  \partial_q {\cal I}_t^{[m_0]}( m,q)  = - \mu_t
+ \frac{ 1 }{4}  \ln \left( \frac{(1+q)^2- (m +  m_0)^2}{(1-q)^2 - (m - m_0)^2}\right)
\label{legendremq}
\end{eqnarray}
while the reciprocal Legendre transformation involves the first derivatives of Eqs \ref{scgfderih} and \ref{scgfderig}
\begin{eqnarray}
{\cal I}_t^{[m_0]}( m,q)  && = hm+gq-   {\cal F} _t^{[m_0]} (h,g)
\label{legendremqreci}
 \\
m && = \partial_h  {\cal F} _t^{[m_0]} (h,g)
=\frac{1+m_0}{2}    \tanh \left[ h + (g+\mu_t)   \right] +\frac{1-m_0}{2} \tanh \left[ h - (g+\mu_t)   \right] 
 = \frac{ \sinh(2h)+ m_0 \sinh(2 [g+\mu_t])}{ \cosh(2h)+  \cosh(2 [g+\mu_t])}
\nonumber \\
q && =    \partial_g {\cal F} _t^{[m_0]} (h,g)=  \frac{1+m_0}{2}   \tanh \left[ h + (g+\mu_t)   \right] - \frac{1-m_0}{2}\tanh \left[ h - (g+\mu_t)   \right]
 = \frac{ m_0 \sinh(2h)+  \sinh(2 [g+\mu_t])}{ \cosh(2h)+  \cosh(2 [g+\mu_t])}
 \nonumber 
\end{eqnarray}


\subsection{ Large deviations properties of the intensive magnetization $m$ alone }

\subsubsection{ Rate function $I _t^{[m_0]} (m) $ for the intensive magnetization $m$ alone }

The rate function $I _t^{[m_0]} (m)  $ for the magnetization $m$ alone corresponds to the optimization
of the joint rate function ${\cal I}_t^{[m_0]}( m,q)   $ over the overlap $q\in [-1,+1] $
 \begin{eqnarray}
I _t^{[m_0]} (m) = \min_{q \in [-1,+1]} \bigg( {\cal I}_t^{[m_0]}( m,q)  \bigg)
\label{rateImcontraction}
\end{eqnarray}
whose properties are described in Appendix \ref{app_malone},
while the Gaussian typical fluctuations of order $\frac{1}{\sqrt N}$ of the magnetization $m$ around its typical value ${\hat m}=\lambda^t m_0$ can be directly obtained from the integration over $q$ of the joint Gaussian approximation of Eq. \ref{taylorgaussapprox}
 \begin{eqnarray}
 p^{[m_0]Gauss}_t( m)   = \int_{-\infty}^{+\infty}  dq p^{[m_0]Gauss}_t( m,q) 
 = \sqrt{ \frac{N}{2 \pi (1-\lambda^{2t} ) } }   e^{ \displaystyle - N \frac{(m-  \lambda^t m_0)^2}{2  (1-\lambda^{2t} )} }
\label{taylorgaussapproxmalone}
\end{eqnarray}
in agreement with the rescaled variance of the magnetization of Eq. \ref{scgfderihg0}.


\subsubsection{ Scaled cumulant generating function $F_t^{[m_0]} (h) $ of the magnetization alone}

The scaled cumulant generating function $F_t^{[m_0]} (h) $ of the magnetization alone
can be directly obtained by plugging the value 
$g=0$ into ${\cal F} _t^{[m_0]} (h,g=0) $ of Eq. \ref{scgf}
 \begin{eqnarray}
F_t^{[m_0]} (h) = {\cal F} _t^{[m_0]} (h,g=0) 
&&  = - \ln (\cosh \mu_t)
  + \frac{1+m_0}{2}  \ln \left(  \cosh \left[ h + \mu_t   \right]  \right)
  + \frac{1-m_0}{2} \ln \left(  \cosh \left[ h - \mu_t   \right]  \right)
\label{scgfhalone}
\end{eqnarray}

For the single variable $m$, the Legendre transformation analogous to Eq. \ref{legendremq} 
involves the derivative of Eq. \ref{rateImcontractionresultderim}
\begin{eqnarray}
F_t^{[m_0]} (h)  && = hm-  I_t^{[m_0]}( m)  
\nonumber \\
h && =    \partial_m  I_t^{[m_0]}( m)   =  \frac{m }{2}  \ln \left(  \frac{ 1+\lambda^t }  { 1-\lambda^t   }\right)
+  \frac{m }{2}  \ln \left(  \frac{1+ m- m_0-  {\mathring q }_t^{[m_0]}(m)}  {1-m -m_0+{\mathring q }_t^{[m_0]}(m)  }\right)
\label{legendremqm}
\end{eqnarray}
while the reciprocal Legendre transformation involves the derivative of Eq. \ref{scgfhalone}
with respect to $h$
\begin{eqnarray}
I_t^{[m_0]}( m)  && = hm-  F_t^{[m_0]} (h) 
\nonumber \\
m && =    \partial_h F_t^{[m_0]} (h)  =  \frac{1+m_0}{2}    \tanh ( h + \mu_t  ) +\frac{1-m_0}{2} \tanh ( h - \mu_t  ) 
= \frac{ \sinh(2h)+ m_0 \sinh(2 \mu_t)}{ \cosh(2h)+  \cosh(2 \mu_t)}
\label{legendremqrecim}
\end{eqnarray}

\subsection{ Discussion}

In summary, the joint statistical properties of the overlap $Q$ and of the magnetization $M$ produced by the forward propagator associated to any given initial condition $\vec \sigma$ of magnetization $M_0$ have been characterized for any time $t$,
both for any finite number $N$ of pixels and in the limit $N \to + \infty$, where the large deviations are 
governed by explicit rate functions and explicit scaled cumulant generating functions.
In the three following sections, the goal will be to analyze the consequences for the backward-generative dynamics
for various initial conditions $P^{ini}$.


\section{ Analysis of the backward dynamics for a single initial condition $\vec \sigma$ }

\label{sec_single}

In this section, we consider the case where the initial condition $P^{ini}(\vec S)$ reduces the single image $\vec \sigma$
\begin{eqnarray}
P^{ini}(\vec S)=\delta_{\vec S,\vec \sigma}
\label{inisingle}
\end{eqnarray}

\subsection{ Direct link with the Doob conditioning of the forward Markov process with a finite horizon $T$}

Plugging the initial condition of Eq. \ref{inisingle}
into 
the backward generated distribution $B^{Gene}_t ( \vec S(t)) $ 
of Eq. \ref{PtbackwardGene} 
 \begin{eqnarray}
B^{Gene}_t ( \vec S(t))
  =  \sum_{\vec S(T) }  P_*(\vec S(T) )
 \frac{ \langle \vec S(T) \vert  {\cal W}^{(T-t)} \vert \vec S(t) \rangle \langle \vec S(t) \vert {\cal W}^t \vert \vec \sigma \rangle
  }{ \langle \vec S(T) \vert {\cal W}^{T} \vert \vec \sigma \rangle }  
\label{PtbackwardGenesigmadef}
\end{eqnarray}
is useful to see the direct links with the theory of conditioned Markov processes introduced by Doob \cite{refDoob,refbookDoob}
(see the recent physics review \cite{refMajumdarOrland} 
and references therein):

(i) when the forward process governed by the Markov generator generator ${\cal W} $ 
starting at the initial condition $\vec \sigma$ is conditioned to end at configuration $\vec S(T) $
at time $T$, the probability $P^{Cond[\vec S(T);\vec \sigma}_t ( \vec S(t)) $
to be at configuration $\vec S(t) $ at the intermediate time $t \in ]0,T[$ is given by the famous bridge formula
 \begin{eqnarray}
P^{Conditioned[\vec S(T);\vec \sigma]}_t ( \vec S(t))
  =  
 \frac{ \langle \vec S(T) \vert  {\cal W}^{(T-t)} \vert \vec S(t) \rangle \langle \vec S(t) \vert {\cal W}^t \vert \vec \sigma \rangle
  }{ \langle \vec S(T) \vert {\cal W}^{T} \vert \vec \sigma \rangle }  
\label{Doobbrige}
\end{eqnarray}

(ii) if the conditioning constraint is instead the whole distribution $P_*(\vec S(T) )$ for the final configuration $\vec S(T) $
at time $T$,
the bridge formula of Eq. \ref{Doobbrige}
is replaced by \cite{refBaudoin,refMultiEnds,us_DoobFirstPassage,us_DoobFirstEncounter}
 \begin{eqnarray}
P^{Conditioned[P_*;\vec \sigma]}_t ( \vec S(t))
  =  
 \sum_{\vec S(T) }  P_*(\vec S(T) ) \frac{ \langle \vec S(T) \vert  {\cal W}^{(T-t)} \vert \vec S(t) \rangle \langle \vec S(t) \vert {\cal W}^t \vert \vec \sigma \rangle
  }{ \langle \vec S(T) \vert {\cal W}^{T} \vert \vec \sigma \rangle }  = B^{Gene}_t ( \vec S(t))
\label{Doobbrigesum}
\end{eqnarray}
that coincides with the backward-generated distribution $B^{Gene}_t ( \vec S(t)) $ 
of Eq. \ref{PtbackwardGene} : the conclusion is thus that the backward dynamics starting with $P_*(\vec S(T) )$ at time $T$
actually generates exactly the same trajectories as the forward process starting at $\vec \sigma$ at time $0$
when the forward process is conditioned to reach its steady state distribution $P_*(.)$
at the finite time $T$.

Let us also stress that for $t=0$, the generated distribution of Eq. \ref{PtbackwardGenesigmadef}
exactly reproduces the single image $\vec \sigma$
 \begin{eqnarray}
B^{Gene}_t ( \vec S(0))
  =  \sum_{\vec S(T) }  P_*(\vec S(T) )
 \frac{ \langle \vec S(T) \vert  {\cal W}^{T} \vert \vec  \sigma \rangle \delta_{\vec S(0),\vec \sigma}
  }{ \langle \vec S(T) \vert {\cal W}^{T} \vert \vec \sigma \rangle }  = \delta_{\vec S(0),\vec \sigma}
\label{PtbackwardGenesigmadeftzero}
\end{eqnarray}


\subsection{ Explicit backward-generated distribution $B^{Gene}_t ( \vec S(t)) $ for any finite $N$ }

In order to compute the backward generated distribution $B^{Gene}_t ( \vec S(t)) $ of Eq. \ref{PtbackwardGenesigmadef} 
and to compare with the forward propagator $\langle \vec S(t) \vert {\cal W}^t \vert \vec \sigma \rangle $,
let us focus on the ratio
 \begin{eqnarray}
\frac{B^{Gene}_t ( \vec S(t))}{\langle \vec S(t) \vert {\cal W}^t \vert \vec \sigma \rangle }
  =  \frac{1}{2^N} 
  \sum_{\vec S(T) }   \frac{ \langle \vec S(T) \vert  {\cal W}^{(T-t)} \vert \vec S(t) \rangle
  }{ \langle \vec S(T) \vert {\cal W}^{T} \vert \vec \sigma \rangle }  
\label{PtbackwardGenesigmaratio}
\end{eqnarray}
where we have replaced the steady state $P_*( \vec S )= \frac{1}{2^N}$ of Eq. \ref{steady}.

The rewriting of the forward propagator $ \langle \vec S(t_2) \vert {\cal W}^{t_2-t_1} \vert \vec S(t_1) \rangle  $ Eq. \ref{Propagatorfactorizedimage}
with the notation of Eq. \ref{mut}
 \begin{eqnarray}
 \langle \vec S(t_2) \vert {\cal W}^{t_2-t_1} \vert \vec S(t_1) \rangle  
&& = \left( \frac{ \sqrt{ 1-\lambda^{2(t_2-t_1)} } }{2} \right)^N
\left( \sqrt{\frac{1+\lambda^{t_2-t_1} }{1- \lambda^{t_2-t_1}} } \right)^{\displaystyle \sum_{n=1}^N S_n(t_2)S_n(t_1) }
\nonumber \\
&& =  \left( \frac{ 1 }{2 \cosh \mu_{(t_2-t_1)} } \right)^N e^{\displaystyle \mu_{(t_2-t_1)} \sum_{n=1}^N S_n(t_2)S_n(t_1)}
\label{Propagatorfactorizedimaget1t2}
\end{eqnarray}
is useful to replace the two propagators on the right handside of Eq. \ref{PtbackwardGenesigmaratio}
to obtain
 \begin{eqnarray}
\frac{B^{Gene}_t ( \vec S(t))}{\langle \vec S(t) \vert {\cal W}^t \vert \vec \sigma \rangle }
&&  =  \frac{1}{2^N} \sum_{\vec S(T) } 
   \frac{ \left( \frac{ 1 }{2 \cosh \mu_{(T-t)} } \right)^N e^{\displaystyle \mu_{(T-t)} \sum_{n=1}^N S_n(T) S_n(t)}
  }{ \left( \frac{ 1 }{2 \cosh \mu_T }\right)^N
e^{\displaystyle \mu_T \sum_{n=1}^N S_n(T) \sigma_n} }  
\nonumber \\
&& =  
\left( \frac{ \cosh \mu_T }{  \cosh \mu_{(T-t)}  } \right)^N
 \frac{1}{2^N} \sum_{\vec S(T) } 
  e^{\displaystyle \mu_{(T-t)} \sum_{n=1}^N S_n(T) S_n(t) - \mu_T \sum_{n=1}^N S_n(T) \sigma_n} 
\label{PtbackwardGenesigmacalcul}
\end{eqnarray}
The last sum corresponds to the generating function of the two overlaps $\sum_{n=1}^N S_n(T) S_n(t) $ 
and $\sum_{n=1}^N S_n(T) \sigma_n $ that reduces to
 \begin{eqnarray}
 \frac{1}{2^N} \sum_{\vec S(T) } 
  e^{\displaystyle \mu_{(T-t)} \sum_{n=1}^N S_n(T) S_n(t) - \mu_T \sum_{n=1}^N S_n(T) \sigma_n} 
&&  
  =  \frac{1}{2^N} \prod_{n=1}^N 
\left[ \sum_{S_n(T) =\pm}  e^{ S_n(T) \left( \mu_{(T-t)}   S_n(t) - \mu_T  \sigma_n \right)} \right]
\nonumber \\
&& = \prod_{n=1}^N 
  \cosh \left[ \mu_{(T-t)}   S_n(t) - \mu_T  \sigma_n \right]
\label{gene2overlap}
\end{eqnarray}

Using that the spins $S_n(t)$ and $\sigma_n$ can take only two values $\pm$ and Eq. \ref{mut}, one can replace
 \begin{eqnarray}
&&  \cosh \left[ \mu_{(T-t)}   S_n(t) - \mu_T  \sigma_n \right]
  = \cosh \mu_{(T-t)}  \cosh \mu_T- S_n(t) \sigma_n\sinh \mu_{(T-t)}  \sinh \mu_T 
\nonumber \\
&& = \cosh \mu_{(T-t)}  \cosh \mu_T \left[ 1 - S_n(t) \sigma_n\tanh \mu_{(T-t)}  \tanh \mu_T \right]
= \cosh \mu_{(T-t)}  \cosh \mu_T \left[ 1 - S_n(t) \sigma_n \lambda^{2T-t} \right]
\nonumber \\
&& 
= \cosh \mu_{(T-t)}  \cosh \mu_T \left[ 1 -  \lambda^{2T-t} \right]^{{\frac{1+S_n(t) \sigma_n}{2} }}
\left[ 1 + \lambda^{2T-t} \right]^{\frac{1-S_n(t) \sigma_n}{2} }
\label{coshspins}
\end{eqnarray}
to obtain that the ratio of Eq. \ref{PtbackwardGenesigmacalcul} 
 \begin{eqnarray}
\frac{B^{Gene}_t ( \vec S(t))}{\langle \vec S(t) \vert {\cal W}^t \vert \vec \sigma \rangle }
&&  =  ( \cosh^2 \mu_T )^2
  \prod_{n=1}^N  \left[ 1 -  \lambda^{2T-t} \right]^{{\frac{1+S_n(t) \sigma_n}{2} }}
\left[ 1 + \lambda^{2T-t} \right]^{\frac{1-S_n(t) \sigma_n}{2} }
\nonumber \\
&& = \frac{1}{(1-\lambda^{2T})^N } 
\left( 1 -  \lambda^{2T-t} \right)^{\frac{N+Q(\vec S(t);\vec \sigma)}{2} }
 \left(  1 +  \lambda^{2T-t}  \right)^{\frac{N-Q(\vec S(t);\vec \sigma)}{2} } 
\label{PtbackwardGenesigmacalculratio}
\end{eqnarray}
depends on the configuration $\vec S(t)$ only via its overlap $Q(\vec S(t);\vec \sigma) $ with the initial configuration $\vec \sigma$
 \begin{eqnarray}
Q(\vec S(t);\vec \sigma) = \sum_{n=1}^N S_n(t) \sigma_n
\label{overlapt}
\end{eqnarray}
Since the forward propagator $\langle \vec S(t) \vert {\cal W}^t \vert \vec \sigma \rangle $ also depends only on 
the overlap $Q(\vec S(t);\vec \sigma) $ via Eq. \ref{Propagatorfactorizedimage}
 \begin{eqnarray}
 \langle \vec S(t) \vert {\cal W}^t \vert \vec \sigma \rangle  
 =    \left( \frac{ 1+\lambda^t  }{2} \right)^{ \frac{N+Q(\vec S(t);\vec \sigma)}{2} }
\left( \frac{ 1- \lambda^t }{2} \right)^{\frac{N-Q(\vec S(t);\vec \sigma)}{2}}
\label{Propagatorfactorizedimagesingle}
\end{eqnarray}
 the final result is that 
the backward-generated distribution $B^{Gene}_t ( \vec S(t)) $ 
 \begin{eqnarray}
B^{Gene}_t ( \vec S(t)) && = 
   \left( \frac{ (1+\lambda^t) (1 -  \lambda^{2T-t}) }{2 (1-\lambda^{2T})} \right)^{\frac{N+Q(\vec S(t);\vec \sigma)}{2}}
\left( \frac{ (1- \lambda^t )(1 +  \lambda^{2T-t})}{2 (1-\lambda^{2T})} \right)^{\frac{N-Q(\vec S(t);\vec \sigma)}{2}}
\nonumber \\
 && =  \left( \frac{ 1+\lambda^t \left[\frac{  1  -  \lambda^{2(T-t)}  }{ 1-\lambda^{2T} } \right] }{2} \right)^{ \frac{N+Q(\vec S(t);\vec \sigma)}{2} }
\left( \frac{ 1- \lambda^t \left[\frac{  1  -  \lambda^{2(T-t)}  }{ 1-\lambda^{2T} } \right]}{2} \right)^{\frac{N-Q(\vec S(t);\vec \sigma)}{2}}
\label{PtbackwardGenesigmaresult}
\end{eqnarray}
has exactly the same form as the forward propagator of Eq. \ref{Propagatorfactorizedimagesingle}
except for the replacement
 \begin{eqnarray}
\lambda^t \to \lambda^t \left[\frac{  1  -  \lambda^{2(T-t)}  }{ 1-\lambda^{2T} } \right]
\label{PtbackwardGenesigmaresultReplacement}
\end{eqnarray}

As a consequence, the probability distribution $b_t^{Gene}(Q) $ of the overlap $Q$ alone   
reduces to the binomial distribution 
 \begin{eqnarray}
b_t^{Gene}(Q) && \equiv \sum_{\vec S(t) } \delta_{Q, \sum_{n=1}^N S_n(t) \sigma_n}B^{Gene}_t ( \vec S(t))
  \nonumber \\
&& = \left( \frac{ 1+\lambda^t \left[\frac{  1  -  \lambda^{2(T-t)}  }{ 1-\lambda^{2T} } \right] }{2} \right)^{ \frac{N+Q(\vec S(t);\vec \sigma)}{2} }
\left( \frac{ 1- \lambda^t \left[\frac{  1  -  \lambda^{2(T-t)}  }{ 1-\lambda^{2T} } \right]}{2} \right)^{\frac{N-Q(\vec S(t);\vec \sigma)}{2}}
\sum_{\vec S(t) } \delta_{Q, \sum_{n=1}^N S_n(t) \sigma_n}
 \nonumber \\
&& =   \left( \frac{ 1+\lambda^t \left[\frac{  1  -  \lambda^{2(T-t)}  }{ 1-\lambda^{2T} } \right] }{2} \right)^{ \frac{N+Q(\vec S(t);\vec \sigma)}{2} }
\left( \frac{ 1- \lambda^t \left[\frac{  1  -  \lambda^{2(T-t)}  }{ 1-\lambda^{2T} } \right]}{2} \right)^{\frac{N-Q(\vec S(t);\vec \sigma)}{2}}
  \frac{ N!}{\frac{N +Q}{2}  !  \frac{N -Q}{2} !}
\label{PtbackwardGenesigmaoverlapbino}
\end{eqnarray}
that has the same form as the forward result of Eq. \ref{ProbaQtalone} up to the replacement of Eq. \ref{PtbackwardGenesigmaresultReplacement}.


\subsection{ Large deviations of the backward-generated distribution of the intensive overlap $q$ for large $N$ }

For large $N$, using the Stirling approximation of Eq. \ref{stirling},
one obtains that the large deviations of the intensive overlap $q=\frac{Q}{N}$ 
in the backward-generated distribution of Eq. \ref{PtbackwardGenesigmaoverlapbino}
 \begin{eqnarray}
b_t^{Gene}(Q=N q) && \oppropto_{N \to + \infty} e^{- N  i_t^{Gene}(q) }
\label{PtbackwardGenesigmaoverlapdinolargedev}
\end{eqnarray}
are governed by the rate function
 \begin{eqnarray}
 i_t^{Gene}(q)  =  \frac{1+q}{2} \ln \left(  \frac{1+q}{1+\lambda^t \left[\frac{  1  -  \lambda^{2(T-t)})  }{ 1-\lambda^{2T}}   \right] } \right)  
  +  \frac{1-q}{2} \ln \left(  \frac{1-q}{ 1-\lambda^t \left[\frac{  1  -  \lambda^{2(T-t)})  }{ 1-\lambda^{2T}}   \right]} \right) 
\label{rateitgeneq}
\end{eqnarray}
that has the same form as the rate function $ i_t (q) $ of Eq. \ref{rateitqdirect}
concerning the forward propagator alone: the only change is that the forward typical value  $\hat q  = \lambda^t$
has been replaced by the backward typical value 
 \begin{eqnarray}
 {\hat q}^{Gene}_t   = \lambda^t \left[\frac{  1  -  \lambda^{2(T-t)})  }{ 1-\lambda^{2T}}   \right]
\label{rateitgeneqtyp}
\end{eqnarray}

\subsection{ Discussion}

In summary, when the initial condition reduces to a single image,
 the backward generative dynamics has very simple properties for any time $t$ 
 in terms of the extensive overlap $Q$ for any finite number $N$ of pixels,
 as well as in terms of the intensive overlap $q=\frac{Q}{N}$ when one focuses on large deviations for large $N \to + \infty$.


\section{ Analysis of the backward dynamics for a mixture of two initial conditions $\vec \sigma^{[a=1,2]}$ }

\label{sec_twoimages}

In this section, we consider the case where the initial condition $P^{ini}(\vec S)$ 
is a mixture of two images $\sigma^{[1]}$ and $\sigma^{[2]}$ 
\begin{eqnarray}
P^{ini}(\vec S)=\frac{1}{2} \delta_{\vec S,\vec \sigma^{[1]}} +\frac{1}{2} \delta_{\vec S,\vec \sigma^{[2]}} 
\label{ini2images}
\end{eqnarray}

\subsection{ Properties of the backward propagator ${\cal B}_{0,T}[ \vec S(0) \vert \vec S(T) ]  $ 
in terms of overlaps}

Plugging the initial condition of Eq. \ref{ini2images} into
the backward propagator ${\cal B}_{0,T}[ \vec S(0) \vert \vec S(T) ]  $ of Eq. \ref{WBackward propagatorfull}
 \begin{eqnarray}
{\cal B}_{0,T}[ \vec S(0) \vert \vec S(T) ] 
&& = \frac{  \langle \vec S(T) \vert {\cal W}^{T} \vert \vec \sigma^{[1]} \rangle \delta_{\vec S(0),\vec \sigma^{[1]}} 
 +\langle \vec S(T) \vert {\cal W}^{T} \vert \vec \sigma^{[2]} \rangle \delta_{\vec S(0),\vec \sigma^{[2]}}   }
 {\langle \vec S(T) \vert {\cal W}^{T} \vert \vec \sigma^{[1]} \rangle 
 +\langle \vec S(T) \vert {\cal W}^{T} \vert \vec \sigma^{[2]} \rangle }
 \nonumber \\
 && = \pi_1(\vec S(T)) \delta_{\vec S(0),\vec \sigma^{[1]}}
 + \pi_2(\vec S(T)) \delta_{\vec S(0),\vec \sigma^{[2]} }
\label{WBackward propagatorfull2im}
\end{eqnarray}
leads to a mixture of the two initial images $\sigma^{[a=1,21]}$ 
where the two weights $\pi_{a=1,2}(\vec S(T)) $ can be rewritten using the forward propagator of Eq. \ref{PtbackwardGenezero}
\begin{eqnarray}
\pi_1(\vec S) && = \frac{  e^{ \displaystyle \mu_T\sum_{n=1}^N S_n\sigma_n^{[1]}}}
 {e^{ \displaystyle \mu_T\sum_{n=1}^N S_n\sigma_n^{[1]} }  + e^{ \displaystyle \mu_T\sum_{n=1}^N S_n\sigma_n^{[2]} } } = \frac{   e^{  \mu_T Q_1}     }{   e^{  \mu_T Q_1}   + e^{  \mu_T Q_2}  }
 \nonumber \\
 \pi_2(\vec S) && = \frac{  e^{ \displaystyle \mu_T\sum_{n=1}^N S_n\sigma_n^{[2]}}}
 {e^{ \displaystyle \mu_T\sum_{n=1}^N S_n\sigma_n^{[1]} }  + e^{ \displaystyle \mu_T\sum_{n=1}^N S_n\sigma_n^{[2]} } } = \frac{   e^{  \mu_T Q_2}     }{   e^{  \mu_T Q_1}   + e^{  \mu_T Q_2}  }
\label{weights2im}
\end{eqnarray}
in terms of the overlaps 
\begin{eqnarray}
Q_1 && \equiv \sum_{n=1}^N S_n \sigma_n^{[1]} 
\nonumber \\
Q_2 && \equiv \sum_{n=1}^N S_n \sigma_n^{[2]} 
\label{Q1Q2}
\end{eqnarray}
of the configuration $\vec S$
with the two given images $\vec \sigma^{[1]} $ and $\vec \sigma^{[2]} $ 
that have a given mutual overlap
\begin{eqnarray}
Q_{12} \equiv \sum_{n=1}^N \sigma_n^{[1]} \sigma_n^{[2]}
\label{Q12}
\end{eqnarray}

In conclusion, for any given initial condition $\vec S(T)$ of the backward-generative dynamics,
one only needs to compute the two overlaps of Eq. \ref{Q1Q2}
to determine the two weights $\pi_{a=1,2}(\vec S(T)) $ that will be generated
for the two images at $t=0$.


\subsection{ Joint probability $P^{[\vec \sigma^{[1]};\vec \sigma^{[2]} ]}(Q_1,Q_2) $ of the two overlaps $Q_1 $ and $Q_2$ with the two given images $[\vec \sigma^{[1]};\vec \sigma^{[2]} ] $}

In order to analyze the joint probability $P^{[\vec \sigma^{[1]};\vec \sigma^{[2]} ]}(Q_1,Q_2) $ 
of the two overlaps of Eq. \ref{Q1Q2}
when the configuration $\vec S$ is drawn with the steady state $P_*( \vec S )= \frac{1}{2^N}$ of Eq. \ref{steady}
\begin{eqnarray}
P^{[\vec \sigma^{[1]};\vec \sigma^{[2]} ]}(Q_1,Q_2) \equiv 
\frac{1}{2^N} \sum_{\vec S }  \delta_{Q_1, \sum_{n=1}^N S_n\sigma_n^{[2]}} 
 \delta_{Q_2,\sum_{n=1}^N S_n\sigma_n^{[1]}}
\label{proba2overlaps12}
\end{eqnarray} 
it is useful to introduce the four populations $(\epsilon_1=\pm,\epsilon_2=\pm)$
\begin{eqnarray}
 N_{\epsilon_1,\epsilon_2} && =\sum_{n=1}^N  \delta_{S_n\sigma_n^{[1]} ,\epsilon_1} \delta_{S_n\sigma_n^{[2]} ,\epsilon_2} 
 =  \frac{ N+ \epsilon_1 Q_1 +\epsilon_2 Q_2   +\epsilon_1 \epsilon_2 Q_{12}}{4}
\label{4populations12}
\end{eqnarray}
with the sum rules
\begin{eqnarray}
\sum_{\epsilon_1=\pm} \sum_{\epsilon_2=\pm}   N_{\epsilon_1,\epsilon_2} && =  N
 \nonumber \\
\sum_{\epsilon_1=\pm} \sum_{\epsilon_2=\pm} \epsilon_1   N_{\epsilon_1,\epsilon_2} && =  Q_1
 \nonumber \\
 \sum_{\epsilon_1=\pm} \sum_{\epsilon_2=\pm} \epsilon_2   N_{\epsilon_1,\epsilon_2}&& =  Q_2
 \nonumber \\
\sum_{\epsilon_1=\pm} \sum_{\epsilon_2=\pm} \epsilon_1 \epsilon_2   N_{\epsilon_1,\epsilon_2} && =  Q_{12}
\label{sumrules12}
\end{eqnarray}
Since the two populations $\frac{1\pm Q_{12}}{2}$ associated to the mutual overlap $Q_{12}$ of Eq. \ref{Q12}
are fixed,
the probability distribution of Eq. \ref{proba2overlaps12}
can be written in terms of the appropriate binomial coefficients
\begin{eqnarray}
P^{[\vec \sigma^{[1]};\vec \sigma^{[2]} ]}(Q_1,Q_2) \equiv \frac{1}{2^N} \times
\frac{ \frac{N +Q_{12}}{2}  !  \frac{N -Q_{12}}{2} !} 
 { \displaystyle \prod_{\epsilon_1=\pm} \prod_{\epsilon_2=\pm}\frac{N+ \epsilon_1 Q_1 +\epsilon_2 Q_2   +\epsilon_1 \epsilon_2 Q_{12}}{4}  !  }
 \equiv P^{[Q_{12} ]}(Q_1,Q_2)
\label{proba2overlaps12bino}
\end{eqnarray}
and depends on the two given images $\vec \sigma^{[1]} $ and $\vec \sigma^{[2]} $ 
only via their mutual overlap $Q_{12}$.


\subsection{ Large deviation properties of the two intensive overlaps $q_1 $ and $q_2$ 
for a given intensive mutual overlap $q_{12}$ }

The large deviations properties of the joint probability distribution $P^{[Q_{12} ]}(Q_1,Q_2) $ of Eq. \ref{proba2overlaps12bino}
in terms of the three intensive overlaps
\begin{eqnarray}
P^{[Q_{12}=N q_{12} ]}(Q_1=N q_1,Q_2=N q_2) \opsimeq_{N \to + \infty} e^{-N {\cal I}^{[q_{12}]}( q_1,q_2)}
\label{proba2overlaps12binolargeN}
\end{eqnarray}
can be analyzed 
as in the analog computation described in the Appendix section \ref{app_stirling}
to obtain the rate function 
 \begin{eqnarray}
 {\cal I}^{[q_{12}]}( q_1,q_2)  
 && =  \ln 2
 - \sum_{\eta=\pm 1}\frac{1+\eta q_{12}}{2}  \ln \left( \frac{1+\eta q_{12}}{2} \right)   
 \nonumber \\ &&
 +   \sum_{\epsilon_1=\pm 1}\sum_{\epsilon_2=\pm 1}  \frac{1+\epsilon_1 q_1 +\epsilon_2 q_2   +\epsilon_1 \epsilon_2 q_{12}}{4}  \ln \left( \frac{1+\epsilon_1 q_1 +\epsilon_2 q_2   +\epsilon_1 \epsilon_2 q_{12}}{4}\right)
 \nonumber \\
 &&
 =  \ln 2 - \frac{1+ q_{12}}{2}  \ln \left( \frac{1+ q_{12}}{2} \right)  - \frac{1- q_{12}}{2}  \ln \left( \frac{1- q_{12}}{2} \right) 
 \nonumber \\ &&
 +    \frac{1+ q_1 + q_2   + q_{12}}{4}  \ln \left( \frac{1+ q_1 + q_2   + q_{12}}{4}\right)
 + \frac{1- q_1 - q_2   + q_{12}}{4}  \ln \left( \frac{1- q_1 - q_2   + q_{12}}{4}\right) 
\nonumber \\ &&
 +    \frac{1+ q_1 - q_2   - q_{12}}{4}  \ln \left( \frac{1+ q_1 - q_2  - q_{12}}{4}\right)
 + \frac{1- q_1 + q_2   - q_{12}}{4}  \ln \left( \frac{1- q_1 + q_2   - q_{12}}{4}\right)
\label{rateIq1q2}
\end{eqnarray}

Since the two weights of Eq. \ref{weights2im} depends only on 
the difference $q_d=(q_1-q_2)$ between the two intensive overlaps
\begin{eqnarray}
\pi_1(\vec S) && = \frac{   e^{  \mu_T Nq_1}     }{   e^{  \mu_T Nq_1}   + e^{  \mu_T Nq_2}  }
= \frac{   e^{  \mu_T N(q_1-q_2) }     }{   e^{  \mu_T N (q_1-q_2)}   + 1  }
\equiv \frac{   e^{  \mu_T N q_d }     }{   e^{  \mu_T N q_d}   + 1  }
 \nonumber \\
 \pi_2(\vec S) && = \frac{   e^{  \mu_T Q_2}     }{   e^{  \mu_T Q_1}   + e^{  \mu_T Q_2}  }
 =\frac{   1    }{   e^{  \mu_T N (q_1-q_2)}   + 1  }
 \equiv \frac{   1    }{   e^{  \mu_T N q_d}   + 1  }
\label{weights2imq12diff}
\end{eqnarray}
it is useful to rewrite the rate function of Eq. \ref{rateIq1q2} 
 \begin{eqnarray}
 {\cal I}^{[q_{12}]}( q_1,q_2)  
  \equiv {\cal J}^{[q_{12}]}( q_s = q_1+q_2, q_d=q_1-q_2) 
\label{rateIq1q1}
\end{eqnarray}
 in terms of the new variables $(q_s = q_1+q_2 ; q_d=q_1-q_2)$
 to obtain their rate function
 \begin{eqnarray}
 {\cal J}^{[q_{12}]}( q_s,q_d)  
 &&  
 =  \ln 2
 - \frac{1+ q_{12}}{2}  \ln \left( \frac{1+ q_{12}}{2} \right)  - \frac{1- q_{12}}{2}  \ln \left( \frac{1- q_{12}}{2} \right)  
 \nonumber \\ &&
 +    \frac{1+ q_s   + q_{12}}{4}  \ln \left( \frac{1+ q_s   + q_{12}}{4}\right)
 + \frac{1- q_s   + q_{12}}{4}  \ln \left( \frac{1- q_s   + q_{12}}{4}\right) 
\nonumber \\ &&
 +    \frac{1+ q_d   - q_{12}}{4}  \ln \left( \frac{1+ q_d  - q_{12}}{4}\right)
 + \frac{1- q_d   - q_{12}}{4}  \ln \left( \frac{1- q_d   - q_{12}}{4}\right)
\label{rateIq1q2J}
\end{eqnarray}

The optimization with respect to $q_s$
\begin{eqnarray}
0= \partial_{q_s} {\cal J}^{[q_{12}]}( q_s,q_d)  
 &&  
 =     \frac{1}{4}  \ln \left( \frac{1+ q_s   + q_{12}}{4}\right)
 - \frac{1}{4}  \ln \left( \frac{1- q_s   + q_{12}}{4}\right) 
 = \frac{1}{4}  \ln \left( \frac{1+ q_s   + q_{12}}{1- q_s   + q_{12}}\right)
\label{rateIq1q2Jderi}
\end{eqnarray}
yields that the vanishing optimal value $q_s^{opt}=0$ can be plugged into Eq. \ref{rateIq1q2J}
to obtain the rate function $J^{[q_{12}]}(q_d) $ of the variable $q_d$ alone
 \begin{eqnarray}
 J^{[q_{12}]}(q_d)  && =   {\cal J}^{[q_{12}]}( q_s^{opt}=0,q_d)    
 \nonumber \\ &&
 = \ln 2
  - \frac{1+ q_{12}}{2}  \ln \left( \frac{1+ q_{12}}{2} \right)  - \frac{1- q_{12}}{2}  \ln \left( \frac{1- q_{12}}{2} \right)
 +    \frac{1   + q_{12}}{2}  \ln \left( \frac{1   + q_{12}}{4}\right)
\nonumber \\ &&
 +    \frac{1+ q_d   - q_{12}}{4}  \ln \left( \frac{1+ q_d  - q_{12}}{4}\right)
 + \frac{1- q_d   - q_{12}}{4}  \ln \left( \frac{1- q_d   - q_{12}}{4}\right)
 \nonumber \\
 && =  - \frac{1- q_{12}}{2} \ln \left( \frac{1- q_{12}}{4} \right) 
    +    \frac{1+ q_d   - q_{12}}{4}  \ln \left( \frac{1+ q_d  - q_{12}}{4}\right)
 + \frac{1- q_d   - q_{12}}{4}  \ln \left( \frac{1- q_d   - q_{12}}{4}\right)
\label{rateIqdiffcontractionres}
\end{eqnarray}
with its derivative
 \begin{eqnarray}
\partial_{q_d} J^{[q_{12}]}(q_d) && =      \frac{1  }{4}  \ln \left( \frac{1+ q_d  - q_{12}}{1- q_d   - q_{12}} \right)
\label{rateIqdiffcontractionresderi}
\end{eqnarray}
that both vanish at the typical value $q_d^{typ}=0$ as expected from the symmetry between $(\pm q_d)$.
The second derivative
 \begin{eqnarray}
\partial_{q_d^2} J^{[q_{12}]}(q_d) \bigg\vert_{q_d=0 }&& =   
   \frac{1  }{4}  \left[ \frac{1}{1+ q_d  - q_{12}}+ \frac{1}{1- q_d   - q_{12}} \right]\vert_{q_d=0 }
   =  \frac{1}{2  (1 - q_{12}) }
\label{rateIqdiffcontractionresderi2}
\end{eqnarray}
 will govern the Gaussian typical fluctuations of order $O\left(\frac{1}{\sqrt N} \right)$ of $q_d$
\begin{eqnarray}
p^{[q_{12} ]}(q_d) \opsimeq_{N \to + \infty} e^{-N \frac{ q_d^2}{4 (1 - q_{12}) }}
\label{gaussqd}
\end{eqnarray}

In conclusion, the ratio of the two weights of Eq. \ref{weights2imq12diff} for the two images
produced by the backward-generative dynamics 
\begin{eqnarray}
\frac{\pi_1(\vec S) }{ \pi_2(\vec S)  }  =   e^{  \mu_T N q_d }  
\label{rarioweights2imq12diff}
\end{eqnarray}
is determined by the difference $q_d=q_1-q_2$
between the two intensive overlaps, whose large deviations for large $N$ are governed
by the rate function $J^{[q_{12}]}(q_d) $ of Eq. \ref{rateIqdiffcontractionres}
and whose Gaussian typical fluctuations of order $O\left(\frac{1}{\sqrt N} \right)$ around 
the origin $q_d=0$ are governed by Eq. \ref{gaussqd}.
As a consequence, the typical scaling of the ratio of Eq. \ref{rarioweights2imq12diff} is $e^{ \pm \sqrt N}$,
i.e. the smallest of the two weights will be of order $e^{ - \sqrt N} $ and thus negligible
with respect to the other complementary weight.
Only the initial conditions that have roughly equal intensive overlaps 
with the two images $q_d =q_1 -q_2\simeq 0$ will produce two finite weights,
but these initial conditions occur only with a probability of order $ \frac{1}{\sqrt N}$.

\subsection{ Discussion}

In summary, when the initial condition reduces to a mixture of two images with equal weights,
 the backward generative dynamics starting at configuration $\vec S(T)$
 will reproduce the two images with unequal weights $\pi_{1,2}(\vec S(T))$ 
 that are determined by the overlaps, and whose statistical properties can be analyzed explicitly
 both for any finite number $N$ of pixels and in the limit of large $N \to + \infty$.


\section{ Analysis when the initial probability $P^{ini}(\vec \sigma)$
is given by the Curie-Weiss model }

\label{sec_curie}

In this section, we focus on the case where the initial probability $P^{ini}(\vec \sigma)$
is given by the Curie-Weiss mean-field ferromagnetic model: 
while the canonical ensemble at inverse temperature $\beta$ has been considered in the previous work \cite{Biroli_largedim}, here we will focus instead on the microcanonical ensemble in order to have an equivalent
of the manifold-hypothesis as discussed in the next subsection.


\subsection{ Equivalent of the manifold-hypothesis for discrete models involving $N$ classical spins $S_n=\pm 1$}

In the field of generative diffusion models, the manifold-hypothesis (see \cite{ManifoldImages,ManifoldHypothesis,Ambrogioni4,Ambrogioni5}
and references therein)
means that the empirical distribution constructed
from the set of interesting images actually lives on a low-dimensional manifold within the global manifold containing all possible images, and the real goal is in fact to learn this low-dimensional manifold.
Indeed in practical applications, one wishes to avoid the perfect memorization
of the initial data and to generate instead new images that are close but nevertheless different from the initial data (see \cite{Ambrogioni2,Biroli_dynamical,Ambrogioni5,forgery,generalization} and references therein).

When translated for our present discrete models involving $N$ classical spins $S_n=\pm 1$ 
where there are $2^N$ possible images,
the analog of the low-dimensional manifold is a subset of images whose size grows less rapidly than $2^N$,
so that the meaningful images of interest become 'rare' for large $N$. 
Let us give two simple examples.

\subsubsection{ Uniform distribution $P^{ini}_{M_F}(\vec \sigma) $ among the configurations $\vec \sigma $ with a given magnetization $M_F$ }

The simplest example is when the configuration $\vec \sigma$ is drawn uniformly
within the set of configurations with a given fixed magnetization $M_F$
\begin{eqnarray}
P^{ini}_{M_F}(\vec \sigma) = \frac{1}{\Omega_N(M_F)}  \delta_{M_F,\sum_n \sigma_n}
\label{iniunifmF}
\end{eqnarray}
whose size $\Omega_N(M_F) $ is given by the binomial coefficient
\begin{eqnarray}
\Omega_N(M_F)  \equiv \sum_{ \vec \sigma}  \delta_{M_F,\sum_n \sigma_n}=  \frac{ N!}{ \left(  \frac{N+M_F}{2}\right)! \left( \frac{N-M_F}{2} \right)!} 
\label{OmegaMF}
\end{eqnarray}
so that the ratio with the total number $2^N$ of possibles images
\begin{eqnarray}
\frac{ \Omega_N(M_F=N m_F) }{2^N}  \opsimeq_{N \to + \infty} e^{- N \left[ \frac{(1+m_F) }{2} \ln (1+m_F) + \frac{(1-m_F)}{2}  \ln (1-m_F)\right] }
\label{OmegaMFlargedev}
\end{eqnarray}
is exponentially small in $N$ as long as the intensive magnetization $m_F=\frac{M_F}{N} \in [-1,+1]$ does not vanish $m_F \ne 0$.

\subsubsection{Curie-Weiss mean-field ferromagnetic model in the microcanonical ensemble where the energy $E_F= - \frac{M_F^2}{2 N} $ is fixed}

In the Curie-Weiss mean-field ferromagnetic model,
the energy $E(\vec \sigma) $ of the configuration $\vec \sigma$
depends only on its magnetization 
\begin{eqnarray}
E(\vec \sigma) = - \frac{1  }{2 N } \left(  \sum_{n=1}^N \sigma_n \right)^2
\label{CurieWeiss}
\end{eqnarray}

So in the microcanonical ensemble where the configuration $\vec \sigma$ is drawn uniformly
among the configurations with a given non-vanishing energy $E_F=- \frac{M_F^2}{2 N}$
that is associated to the two opposite magnetizations $(\pm M_F)$, the distribution
reduces to the mixture of the two fixed-magnetization distributions $P^{ini}_{\pm M_F}(\vec \sigma) $ of Eq. \ref{iniunifmF}
\begin{eqnarray}
P^{ini}_{E_F=- \frac{M_F^2}{2 N}}(\vec \sigma) = \frac{1}{2} \left( P^{ini}_{M_F}(\vec \sigma)  +P^{ini}_{-M_F}(\vec \sigma) \right)
= \frac{1}{2 \Omega_N(M_F)} \left(  \delta_{M_F,\sum_n \sigma_n} +\delta_{-M_F,\sum_n \sigma_n} \right)
\label{inicurie}
\end{eqnarray}


\subsection{ Properties of the forward dynamics for the magnetization $M$}

The forward dynamics in the space of configurations $\vec S$
can be projected onto the forward dynamics for the magnetization $M$ alone
 \begin{eqnarray}
p_t(M ) && = \sum_{M_0} p_t(M \vert M_0) p^{ini}(M_0) 
\label{forwardsolum0}
\end{eqnarray}
where the initial distribution $p^{ini}(M_0) $ reduces to a single delta function for the case of Eq. \ref{iniunifmF}
\begin{eqnarray}
p^{ini}_{M_F}(M_0) =   \delta_{M_0,M_F}
\label{iniunifmFM0}
\end{eqnarray}
or to the mixture of two delta functions for the case of Eq. \ref{inicurie}
\begin{eqnarray}
p^{ini}_{E_F=- \frac{M_F^2}{2 N}}(M_0) 
= \frac{1}{2 } \left(  \delta_{M_0,M_F} +\delta_{M_0,-M_F} \right)
\label{inicurieM0}
\end{eqnarray}

The forward propagator $p_t(M \vert M_0) $ governing Eq. \ref{forwardsolum0}
corresponds to the sum over the overlap $Q$
 of the joint distribution $P^{[M_0]}_t (M,Q) $ analyzed in Eqs \ref{JointMQ} and \ref{JointMQbino}
 \begin{eqnarray}
p_t(M \vert M_0) && =\sum_Q P^{[M_0]}_t (M,Q) 
\nonumber \\
&& = \sum_Q  \left( \frac{ 1+\lambda^t  }{2} \right)^{ \frac{N+Q}{2} }
\left( \frac{ 1- \lambda^t }{2} \right)^{\frac{N-Q}{2} }
\frac{ \frac{N +M_0}{2}  !  \frac{N -M_0}{2} !} 
 {\frac{N +  M_0 + M +  Q}{4}  !  \frac{ N+ M_0 - M   -  Q }{4}!\frac{N- M_0 + M   -  Q}{4}!  \frac{ N- M_0 - M   +  Q }{4}!}
\label{forwardpropagatormm0}
\end{eqnarray}
so that its large deviations properties are governed by the rate function $I_t^{[m_0]}(m) $ introduced in Eq. \ref{rateImcontraction}
 \begin{eqnarray}
p_t(M=Nm \vert M_0=N m_0)  \oppropto_{N \to + \infty} e^{-N I_t^{[m_0]}(m)}
\label{forwardpropagatormm0largedev}
\end{eqnarray}


\subsection{ Properties of the backward dynamics for the magnetization $M$}

Since the uniform steady state $P_*(\vec S(T))=\frac{1}{2^N}$ in the spin configuration space
translates into the binomial distribution $p_*(M_T) $ for the global magnetization $M_T$
\begin{eqnarray}
p_*(M_T) = \frac{1}{2^N}\sum_{\vec S(T)}  \delta_{M_T,\sum_n S_n(T)}
= 2^{-N} \frac{ N!}{ \left(  \frac{N+M_T}{2}\right)! \left( \frac{N-M_T}{2} \right)!} 
\label{pmtsteadybino}
\end{eqnarray}
the backward-generated distribution $b^{Gene}_t (M)$ of the magnetization $M$ at time $t$ 
translated from Eq. \ref{PtbackwardGene} reads
 \begin{eqnarray}
b^{Gene}_t (M)
&& = p_t(M )  \sum_{M_T }  p_*(M_T ) \frac{ p_{T-t}( M_T \vert M )  }
{  p_T(M_T ) }  
\nonumber \\
&& =  p_t(M ) \sum_{M_T } 2^{-N} \frac{ N!}{ \left(  \frac{N+M_T}{2}\right)! \left( \frac{N-M_T}{2} \right)!} 
 \times  \frac{ p_{T-t}( M_T \vert M )   }{  p_T(M_T ) }  
\label{PtbackwardGenemagne}
\end{eqnarray}
with the simplification for $t=0$
 \begin{eqnarray}
b^{Gene}_{t=0} (M_0)
&& = p^{ini}(M_0) \sum_{M_T } 2^{-N} \frac{ N!}{ \left(  \frac{N+M_T}{2}\right)! \left( \frac{N-M_T}{2} \right)!} 
 \times  \frac{ p_{T}( M_T \vert M_0 )   }{  \sum_{M_0'} p_T(M_T \vert M_0') p^{ini}(M_0') }  
\label{PtbackwardGenemagnet0}
\end{eqnarray}
and with the large deviations of the steady distribution $p_*(M_T) $ of Eq. \ref{pmtsteadybino}
 \begin{eqnarray}
 p_*(M_T) = 2^{-N} \frac{ N!}{ \left(  \frac{N+M_T}{2}\right)! \left( \frac{N-M_T}{2} \right)!} 
  \oppropto_{N \to + \infty} e^{  - N \left[ \frac{1+ m_T}{2}  \ln \left( 1+m_T \right) 
  + \frac{1- m_T}{2}  \ln \left( 1- m_T \right)\right] } 
\label{binolargedev}
\end{eqnarray}


\subsubsection{ Application to the fixed-magnetization ensemble $p^{ini}_{M_F}(M_0) =   \delta_{M_0,M_F} $}

For the fixed-magnetization case of Eq. \ref{iniunifmFM0},
the forward solution $p_t(M ) $ of Eq. \ref{forwardsolum0} reduces to the forward propagator $p_t(M \vert M_F) $
whose large deviations properties of Eq. \ref{forwardpropagatormm0largedev}
are governed by the rate function $I_t^{[m_F]}(m) $
 \begin{eqnarray}
p_t(M )  =  p_t(M \vert M_F)  \oppropto_{N \to + \infty} e^{-N I_t^{[m_F]}(m)}
\label{forwardsolum0MF}
\end{eqnarray}

The backward-generated distribution $b^{Gene}_t (M)$ of Eq. \ref{PtbackwardGenemagne} becomes
 \begin{eqnarray}
b^{Gene}_t (M)
&& = p_t(M \vert M_F)  \sum_{M_T }  p_*(M_T ) \frac{ p_{T-t}( M_T \vert M )  }
{  p_T(M_T \vert M_F) }  
\nonumber \\
&& =   p_t(M \vert M_F) \sum_{M_T } 2^{-N} \frac{ N!}{ \left(  \frac{N+M_T}{2}\right)! \left( \frac{N-M_T}{2} \right)!} 
 \times  \frac{ p_{T-t}( M_T \vert M )   }{  p_T(M_T \vert M_F) }  
\label{PtbackwardGenemagneMF}
\end{eqnarray}
with the large deviations behavior 
 \begin{eqnarray}
 b^{Gene}_t (M)
&& \opsimeq_{N \to + \infty}    e^{-N I_t^{[m_F]}(m)}
\int_{-1}^{+1} dm_T
e^{  - N \left[ \frac{1+ m_T}{2}  \ln \left( 1+m_T \right) 
  + \frac{1- m_T}{2}  \ln \left( 1- m_T \right) -I_{T-t}^{[m]}(m_T)+ I_T^{[m_F]}(m_T)\right] } 
  \nonumber \\
&&  \opsimeq_{N \to + \infty}    e^{ \displaystyle -N \left( I_t^{[m_F]}(m) +J_t^{[m_F]}(m)\right)} 
\label{PtbackwardGenemagnelargedev}
\end{eqnarray}
where the supplementary contribution $J_t^{[m_F]}(m) $ with respect to $I_t^{[m_F]}(m) $ governing the forward solution
of Eq. \ref{forwardsolum0MF}
 \begin{eqnarray}
J_t^{[m_F]}(m) \equiv \min_{m_T\in [-1,1]} \left[ \frac{1+ m_T}{2}  \ln \left( 1+m_T \right) 
  + \frac{1- m_T}{2}  \ln \left( 1- m_T \right) -I_{T-t}^{[m]}(m_T)+ I_T^{[m_F]}(m_T)\right]
\label{Jauxiliaire}
\end{eqnarray}
requires the optimization over $m_T \in [-1,1]$ of the function in the square parenthesis
that involves the difference between the two rate functions $I_{T-t}^{[m]}(m_T) $ and $I_T^{[m_F]}(m_T) $.


\subsubsection{ Application to the Curie-Weiss microcanonical ensemble $p^{ini}_{E_F=- \frac{M_F^2}{2 N}}(M_0) 
= \frac{1}{2 } \left(  \delta_{M_0,M_F} +\delta_{M_0,-M_F} \right) $}

For the Curie-Weiss microcanonical ensemble of Eq. \ref{inicurieM0},
the forward solution $p_t(M ) $ of Eq. \ref{forwardsolum0} reduces to the mixture of two forward propagators
whose large deviations properties of Eq. \ref{forwardpropagatormm0largedev}
are governed by the two rate functions $I_t^{[\pm m_F]}(m) $
 \begin{eqnarray}
p_t(M )  && = \frac{p_t(M \vert M_F)  +p_t(M \vert - M_F)}{2}  
 \oppropto_{N \to + \infty} 
\frac{e^{-N I_t^{[m_F]}(m)}  +e^{-N I_t^{[-m_F]}(m)}}{2} 
\nonumber \\
&& \oppropto_{N \to + \infty} e^{-N \min \left[ I_t^{[m_F]}(m) ; I_t^{[-m_F]}(m) \right]}
 \oppropto_{N \to + \infty} e^{-N  I_t^{[m_F {\rm sgn}(m)]}(m) }
\label{forwardsolum0CW}
\end{eqnarray}
where the minimum between the two rate functions $I_t^{[\pm m_F]}(m) $
is obtained when the two magnetizations have the same sign.

The backward-generated distribution $b^{Gene}_t (M)$ of Eq. \ref{PtbackwardGenemagne} becomes
 \begin{eqnarray}
b^{Gene}_t (M)
&& = \bigg( p_t(M \vert M_F)  +p_t(M \vert - M_F)\bigg)  \sum_{M_T }  p_*(M_T ) \frac{ p_{T-t}( M_T \vert M )  }
{   p_T(M_T \vert M_F)  +p_T(M_T \vert - M_F)  }  
\nonumber \\
&& = \bigg( p_t(M \vert M_F)  +p_t(M \vert - M_F)\bigg)  \sum_{M_T } 2^{-N} \frac{ N!}{ \left(  \frac{N+M_T}{2}\right)! \left( \frac{N-M_T}{2} \right)!} 
 \times  \frac{ p_{T-t}( M_T \vert M )   }{  p_T(M_T \vert M_F)  +p_T(M_T \vert - M_F) }  
\label{PtbackwardGenemagneCW}
\end{eqnarray}
with the large deviations behavior for the intensive magnetization $m$
 \begin{eqnarray}
 b^{Gene}_t (M=Nm)
&& \opsimeq_{N \to + \infty}  e^{-N  I_t^{[m_F {\rm sgn}(m)]}(m) }
\int_{-1}^{+1} dm_T
 e^{  - N \left[ \frac{1+ m_T}{2}  \ln \left( 1+m_T \right) 
  + \frac{1- m_T}{2}  \ln \left( 1- m_T \right) -I_{T-t}^{[m]}(m_T)+ I_T^{[m_F{\rm sgn}(m_T)]}(m_T) \right] } 
  \nonumber \\
&&  \opsimeq_{N \to + \infty}    e^{ \displaystyle -N \left( I_t^{[m_F {\rm sgn}(m)]}(m)  +K_t^{[m_F]}(m)\right)} 
\label{PtbackwardGeneCVlargedev}
\end{eqnarray}
where the supplementary contribution $K_t^{[m_F]}(m) $ with respect to $I_t^{[m_F {\rm sgn}(m)]}(m)$ governing the forward solution
of Eq. \ref{forwardsolum0CW}
 \begin{eqnarray}
K_t^{[m_F]}(m) \equiv \min_{m_T\in [-1,1]} \left[ \frac{1+ m_T}{2}  \ln \left( 1+m_T \right) 
  + \frac{1- m_T}{2}  \ln \left( 1- m_T \right) -I_{T-t}^{[m]}(m_T)+ I_T^{[m_F{\rm sgn}(m_T)]}(m_T)\right]
  \ \ \ \ \ \ 
\label{Kauxiliaire}
\end{eqnarray}
requires the optimization over $m_T \in [-1,1]$ of the function in the square parenthesis
that involves the difference between the two rate functions $I_{T-t}^{[m]}(m_T) $ and $I_T^{[m_F{\rm sgn}(m_T)]}(m_T) $.

 
\section{ Conclusions  }

\label{sec_conclusions}

 In this paper, we have advertized discrete-time Markov spin models for image generation, 
and we have stressed the similarities and the differences with respect to the standard
generative continuous diffusion models. 
We have focused on the simplest case when each pixel $n=1,..,N$ can take only two values $S_n=\pm 1$,
but it would be of course interesting in the future to go beyond these black-and-white images and to explore the case of colored images where each pixel can be in $C>2$ different colors or more generally to adapt to the particular encoding of the data one is interested in.

We have first discussed the forward trivial dynamics, where the $N$ pixels evolve towards noise independently. 
The key observation is that the forward propagator of Eq. \ref{Propagatorfactorizedimage} 
depends on the initial and final configurations of the $N$ spins only via their extensive overlap $Q$ that simply counts the number of spins that have the same value or not in the two configurations. 
In statistical physics models involving a large number $N$ of spins, extensive global variables
corresponding to sums over the $N$ spins play of course a major role, in particular the magnetization $M$ 
of a single configuration which is the standard order parameter for ferromagnetic systems, 
while the overlap $Q$ between two configurations is also very familiar from the field of spin-glasses and glassy systems,
and more generally whenever the type of order is not obvious, as in the present generative models where the interesting images
can be anything.
We have explained how the simple form of the forward propagator 
leads to an explicit joint probability distribution of the overlap $Q$ and of the magnetization $M$
at time $t$ for any finite number $N$ of pixels, from which it is straightforward to extract the large deviations properties of the corresponding intensive variables $q=\frac{Q}{N}$ and $m=\frac{M}{N}$ when the number $N$ of pixels becomes large $N \to + \infty$.

We have then turned to the backward-generative dynamics
that is based on the mathematical possibility of reversing the time-arrow in Markov processes
and that is able to 'create order out of noise', i.e. to drive the system from high entropy towards low entropy. 
Indeed, in the present spins models of $N$ spins, 
the 'noise' distribution $P_*(\vec S)$ taken as the initial condition of the backward dynamics
is simply the maximum-entropy distribution where the $2^N$ possible images have the same probability $2^{-N}$,
while the target of the backward dynamics is the 'rare' subset where the meaningful images of interest are 
actually localized, in agreement with the manifold-hypothesis concerning continuous-space diffusion models.
To be concrete, we have focused on three very simple initial conditions:

(i) when the initial condition reduces to a single image,
we have obtained that
 the backward generative dynamics has explicit properties for any time $t$ 
 in terms of the extensive overlap $Q$ for any finite number $N$ of pixels,
 as well as in terms of the intensive overlap $q=\frac{Q}{N}$ in the limit $N \to + \infty$.
We have also explained the direct link with the Doob conditioning of Markov processes.

(ii) when the initial condition reduces to a mixture of two images with equal weights,
 the backward generative dynamics starting at configuration $\vec S(T)$
 will reproduce the two images with unequal weights $\pi_{1,2}(\vec S(T))$ 
 that are determined by the overlaps between $\vec S(T) $ and the two target-images, 
 and whose statistical properties can be analyzed explicitly
 both for any finite number $N$ of pixels and in the limit of large $N \to + \infty$.

(iii) when the initial condition is the uniform distribution on the subset of constant magnetization $M_F$,
or is given by the uniform distribution on the subset of constant opposite magnetizations $(\pm M_F)$ that corresponds to the microcanonical ensemble of the Curie-Weiss mean-field ferromagnetic model, the forward and the backward dynamics for the $N$ spins can be analyzed via their projections for the global magnetization $M$, with the corresponding large deviations properties for the intensive magnetization $m=\frac{M}{N}$ in the limit $N \to + \infty$.

In conclusion, we hope that, beyond the simple examples described in the present paper, the theory of large deviations 
 will be helpful in the future to clarify many other interesting issues raised by the impressive efficiency of generative Markov models.



\appendix

\section{ Rate function ${\cal I}_t^{[m_0]} (m,q) $ for the intensive magnetization $m$ and the intensive overlap $q$}

\subsection{ Explicit calculation from the joint probability distribution $P^{[M_0=N m_0]}_t (M=N m,Q=N q) $ for finite $N$}

\label{app_stirling}

To analyze the behavior of the joint probability distribution $P^{[M_0=N m_0]}_t (M=N m,Q=N q) $ 
of Eq. \ref{JointMQbinorescal} when the finite number $N$ of pixels becomes large $N \to + \infty$, one only needs
to use the Stirling formula of Eq. \ref{stirling} as follows.
Let us evaluate the logarithmic contribution from the two factorials in the numerator
 \begin{eqnarray}
&& \ln \left[  \left( N \frac{1+m_0}{2} \right) !  \left( N \frac{1-m_0}{2} \right) ! \right] = \sum_{\eta=\pm 1} \ln \left[ \left( N \frac{1+\eta m_0}{2} \right)! \right]
\nonumber \\
&&  \opsimeq_{N \to + \infty} 
\sum_{\eta=\pm 1} \left[ \frac{1+\eta m_0}{2} N   \ln(N) 
+ N \frac{1+\eta m_0}{2}  \ln \left( \frac{1+\eta m_0}{2} \right)
- N \frac{1+\eta m_0}{2}  
+  \frac{1}{2} \ln (2 \pi N)
+\frac{1}{2} \ln \left(  \frac{1+\eta m_0}{2} \right)
+ O\left( \frac{1}{N} \right)\right]
\nonumber \\
&&  \opsimeq_{N \to + \infty} 
 N  \ln(N) 
+ N \sum_{\eta=\pm 1}\frac{1+\eta m_0}{2}  \ln \left( \frac{1+\eta m_0}{2} \right)
- N  
   +   \ln (2 \pi N)
+ \frac{1}{2} \sum_{\eta=\pm 1} \ln \left(  \frac{1+\eta m_0}{2} \right)
+ O\left( \frac{1}{N} \right)
\label{stirling1}
\end{eqnarray}
as well as the logarithmic contribution from the four factorials in the denominator 
 \begin{eqnarray}
&&   \ln \left[\prod_{\epsilon=\pm} \prod_{\epsilon_0=\pm} \left( N \frac{1+\epsilon m + \epsilon_0 m_0+\epsilon \epsilon_0 q}{4} \right)  ! \right]
= \sum_{\epsilon=\pm 1}\sum_{\epsilon_0=\pm 1} 
\ln \left[ N \left( \frac{1+\epsilon m + \epsilon_0 m_0+\epsilon \epsilon_0 q}{4} \right)! \right]
\nonumber \\
&&  \opsimeq_{N \to + \infty} 
 \sum_{\epsilon=\pm 1}\sum_{\epsilon_0=\pm 1} 
 \bigg[
\frac{1+\epsilon m + \epsilon_0 m_0+\epsilon \epsilon_0 q}{4} N   \ln N 
+ N \frac{1+\epsilon m + \epsilon_0 m_0+\epsilon \epsilon_0 q}{4}  \ln \left( \frac{1+\epsilon m + \epsilon_0 m_0+\epsilon \epsilon_0 q}{4}\right) 
\nonumber \\
&& - N \frac{1+\epsilon m + \epsilon_0 m_0+\epsilon \epsilon_0 q}{4}  
+  \frac{1}{2} \ln (2 \pi N)
+ \frac{1}{2} \ln \left( \frac{1+\epsilon m + \epsilon_0 m_0+\epsilon \epsilon_0 q}{4}\right)
 + O\left( \frac{1}{N} \right)
 \bigg]
\nonumber \\
&&  \opsimeq_{N \to + \infty} 
 N   \ln N 
+ N  \sum_{\epsilon=\pm 1}\sum_{\epsilon_0=\pm 1}  \frac{1+\epsilon m + \epsilon_0 m_0+\epsilon \epsilon_0 q}{4}  \ln \left( \frac{1+\epsilon m + \epsilon_0 m_0+\epsilon \epsilon_0 q}{4}\right) 
\nonumber \\
&& - N 
+ 2 \ln (2 \pi N)
+ \frac{1}{2}  \sum_{\epsilon=\pm 1}\sum_{\epsilon_0=\pm 1} \ln \left( \frac{1+\epsilon m + \epsilon_0 m_0+\epsilon \epsilon_0 q}{4}\right)
+ O\left( \frac{1}{N} \right)
\label{stirling2}
\end{eqnarray}

The difference between Eq. \ref{stirling1} and Eq. \ref{stirling2}
needed for Eq. \ref{JointMQbinorescal} reduces to
 \begin{eqnarray}
&& \ln \left[ \frac{ \left( N \frac{1+m_0}{2} \right) !  \left( N \frac{1-m_0}{2} \right) ! } 
 { \displaystyle \prod_{\epsilon=\pm} \prod_{\epsilon_0=\pm} \left( N \frac{1+\epsilon m + \epsilon_0 m_0+\epsilon \epsilon_0 q}{4} \right)  ! } \right]
=
\sum_{\eta=\pm 1} \ln \left[ \left( N \frac{1+\eta m_0}{2} \right)! \right]
-  \sum_{\epsilon=\pm 1}\sum_{\epsilon_0=\pm 1}  \ln \left[ N \left( \frac{1+\epsilon m + \epsilon_0 m_0+\epsilon \epsilon_0 q}{4} \right)! \right]
\nonumber \\
&&  \opsimeq_{N \to + \infty} 
 N \left[ \sum_{\eta=\pm 1}\frac{1+\eta m_0}{2}  \ln \left( \frac{1+\eta m_0}{2} \right)   
-   \sum_{\epsilon=\pm 1}\sum_{\epsilon_0=\pm 1}  \frac{1+\epsilon m + \epsilon_0 m_0+\epsilon \epsilon_0 q}{4}  \ln \left( \frac{1+\epsilon m + \epsilon_0 m_0+\epsilon \epsilon_0 q}{4}\right) \right]
\nonumber \\
&& 
-  \ln ( 2 \pi N ) 
+ \frac{1}{2} \left[ \sum_{\eta=\pm 1} \ln \left(  \frac{1+\eta m_0}{2} \right)
-   \sum_{\epsilon=\pm 1}\sum_{\epsilon_0=\pm 1} \ln \left( \frac{1+\epsilon m + \epsilon_0 m_0+\epsilon \epsilon_0 q}{4}\right) \right]
+ O\left( \frac{1}{N} \right)
\label{stirlingdiff}
\end{eqnarray}
that
 leads to the explicit rate function ${\cal I}_t^{[m_0]} (m,q) $ written in Eq. \ref{rateImq} of the main text.


\subsection{ Alternative expression of the rate function ${\cal I}_t^{[m_0]} (m,q) $ of Eq. \ref{rateImq}}

\label{app_jointrate}

 One can use the first derivatives of Eqs \ref{rateImqderim} and \ref{rateImqderiq}
to rewrite the rate function of Eq. \ref{rateImq}
as
 \begin{eqnarray}
  {\cal I}_t^{[m_0]}( m,q)  && =  m \partial_m  {\cal I}_t^{[m_0]}( m,q)  + q \partial_q  {\cal I}_t^{[m_0]}( m,q)  
- \frac{1}{2}  \ln \left( \frac{ 1- \lambda^{2t} }{4} \right)
- \sum_{\eta=\pm 1}\frac{1+\eta m_0}{2}  \ln \left( \frac{1+\eta m_0}{2} \right)   
\nonumber \\
&& +   \sum_{\epsilon_0=\pm 1}  \frac{1
 + \epsilon_0 m_0}{4} \sum_{\epsilon=\pm 1} \ln \left( \frac{1+\epsilon m + \epsilon_0 m_0+\epsilon \epsilon_0 q}{4}\right)
 \nonumber \\
&& =  m \partial_m  {\cal I}_t^{[m_0]}( m,q)  + q \partial_q  {\cal I}_t^{[m_0]}( m,q)  
- \frac{1}{2}  \ln \left( \frac{ 1- \lambda^{2t} }{4} \right)
- \sum_{\epsilon_0=\pm 1}\frac{1+\epsilon_0 m_0}{2}  \ln \left( \frac{1+\epsilon_0 m_0}{2} \right)   
\nonumber \\
&& +   \sum_{\epsilon_0=\pm 1}  \frac{1 + \epsilon_0 m_0}{4} 
 \ln \left( \frac{(1+ \epsilon_0 m_0)^2-( m + \epsilon_0 q)^2}{4^2}\right) 
\label{rateImqavecderi}
\end{eqnarray}

Since this rate function ${\cal I}_t^{[m_0]}( m,q) $
vanishes at the typical value $(\hat m,\hat q)$ of Eq. \ref{bothderivanisheq}
 \begin{eqnarray}
&&  {\cal I}_t^{[m_0]}( \hat m = \lambda^t m_0, \hat q = \lambda^t)   = 
- \frac{1}{2}  \ln \left( \frac{ 1- \lambda^{2t} }{4} \right)
- \sum_{\epsilon_0=\pm 1}\frac{1+\epsilon_0 m_0}{2}  \ln \left( \frac{1+\epsilon_0 m_0}{2} \right)   
 +   \sum_{\epsilon_0=\pm 1}  \frac{1 + \epsilon_0 m_0}{4} 
 \ln \left( \frac{(1+ \epsilon_0 m_0)^2( 1- \lambda^{2t}) }{4^2}\right) 
 \nonumber \\
&& = - \frac{1}{2}  \ln \left( \frac{ 1- \lambda^{2t} }{4} \right)
- \sum_{\epsilon_0=\pm 1}\frac{1+\epsilon_0 m_0}{2}  \ln \left( \frac{1+\epsilon_0 m_0}{2} \right)   
 +   \sum_{\epsilon_0=\pm 1}  \frac{1 + \epsilon_0 m_0}{4} 
\left[ 2 \ln \left( \frac{1+ \epsilon_0 m_0 }{2}\right) 
+ \ln \left( \frac{ 1- \lambda^{2t} }{4}\right)
\right] =0
\label{rateImqzerotyp}
\end{eqnarray}
it is convenient to use this equation to replace
 \begin{eqnarray}
- \frac{1}{2}  \ln \left( \frac{ 1- \lambda^{2t} }{4} \right)
- \sum_{\epsilon_0=\pm 1}\frac{1+\epsilon_0 m_0}{2}  \ln \left( \frac{1+\epsilon_0 m_0}{2} \right)   
=
 -   \sum_{\epsilon_0=\pm 1}  \frac{1 + \epsilon_0 m_0}{4} 
 \ln \left( \frac{(1+ \epsilon_0 m_0)^2( 1- \lambda^{2t}) }{4^2}\right) 
 \label{rateImqzerotypreplace}
\end{eqnarray}
into Eq. \ref{rateImqavecderi} to obtain the following alternative expression for the rate function
 \begin{eqnarray}
  {\cal I}_t^{[m_0]}( m,q)  && =  m \partial_m  {\cal I}_t^{[m_0]}( m,q)  + q \partial_q  {\cal I}_t^{[m_0]}( m,q)    
 +   \sum_{\epsilon_0=\pm 1}  \frac{1 + \epsilon_0 m_0}{4} 
 \ln \left( \frac{(1+ \epsilon_0 m_0)^2-( m + \epsilon_0 q)^2}{(1+ \epsilon_0 m_0)^2( 1- \lambda^{2t})}\right) 
\label{rateImqavecderiusingtypt}
\end{eqnarray}
whose vanishing at the typical value $( \hat m = \lambda^t m_0, \hat q = \lambda^t)$ of the pair $(m,q)$ is now obvious.


\subsection{ Recovering the large deviations properties of the intensive overlap $q$  alone }

\label{app_recoveringq}

\subsubsection{ Recovering the explicit rate function $i _t (q) $ of Eq. \ref{rateitqdirect} for the intensive overlap $q$ alone }

The rate function $i_t^{[m_0]} (q) $ for the overlap $q$ alone corresponds to the optimization
of the joint rate function ${\cal I}_t^{[m_0]}( m,q)   $ over the magnetization $m$
 \begin{eqnarray}
i_t^{[m_0]} (q) = \min_{m} \bigg( {\cal I}_t^{[m_0]}( m,q)  \bigg)
\label{rateIqcontraction}
\end{eqnarray}

The derivative of Eq. \ref{rateImqderim}
 \begin{eqnarray}
0 = \partial_m  {\cal I}_t^{[m_0]}( m,q)  
 = \frac{1 }{4}  \ln \left( \frac{(1+ m)^2 - (  m_0+  q)^2}{(1-m)^2 -(m_0-q)^2}\right)
\label{rateImqderimvanish}
\end{eqnarray}
leads to the optimal magnetization $m^{opt} $ as a function of any given overlap $q$
 \begin{eqnarray}
m^{opt} =  m_0  q
\label{rateImqderimvanishmopt}
\end{eqnarray}
that can be plugged into the rate function $ {\cal I}_t^{[m_0]}( m,q) $ of Eq. \ref{rateImqavecderiusingtypt} 
 \begin{eqnarray}
 i_t^{[m_0]} (q) && =  {\cal I}_t^{[m_0]}( m^{opt}=m_0 q,q)  
\nonumber \\
&& =  
 q \left[ - \frac{1}{2} \ln \left( \frac{ 1+\lambda^t  }{1-\lambda^t } \right)
+ \frac{ 1 }{4}  \ln \left( \frac{(1+q)^2 (1-  m_0^2)}{(1-q)^2(1-  m_0^2)}\right)  \right]  
 +   \sum_{\epsilon_0=\pm 1}  \frac{1 + \epsilon_0 m_0}{4} 
 \ln \left( \frac{(1+ \epsilon_0 m_0)^2(1-q^2)}{(1+ \epsilon_0 m_0)^2( 1- \lambda^{2t})}\right) 
 \nonumber \\
&& =  
 \frac{ q }{2}  \ln \left( \frac{(1+q) (1-\lambda^t ) }{ (1+\lambda^t ) (1-q)}\right)  
 +     \frac{1 }{2} 
 \ln \left( \frac{(1-q^2)}{( 1- \lambda^{2t})}\right) 
  \nonumber \\
&& =  
 \frac{ 1+q }{2}  \ln \left( \frac{1+q  }{ 1+\lambda^t }\right)  
+  \frac{ 1-q }{2}  \ln \left( \frac{1-q  }{ 1-\lambda^t }\right)  \equiv  i_t (q)
\label{rateImqavecderiusingtyptcontractionmopt}
\end{eqnarray}
to recover the rate function $ i_t (q)$ of Eq. \ref{rateitqdirect} as it should.

\subsubsection{  Scaled cumulant generating function $f_t(g) $ for the overlap $q$ alone}

The scaled cumulant generating function $f_t(g) $ for the overlap $q$ alone
can be directly obtained by plugging the value 
$h=0$ into ${\cal F} _t^{[m_0]} (h=0,g) $ of Eq. \ref{scgf}
 \begin{eqnarray}
 {\cal F} _t^{[m_0]} (h=0,g) 
   = \ln \left(  \cosh \left[ g+\mu_t   \right]  \right) - \ln (\cosh \mu_t) \equiv f_t(g)
 \label{scgfgalone}
\end{eqnarray}

For the single variable $q$, the Legendre transformation 
between the rate function $i_t(q)$ and the scaled cumulant generating function $f_t(g) $
analogous to Eq. \ref{legendremq} involves the derivative of Eq. \ref{rateImqavecderiusingtyptcontractionmopt}
with respect to $q$
\begin{eqnarray}
f_t(g) && = gq-  i_t( q)  
\nonumber \\
g && =   \frac{ d i_t (q) }{dq}   =\frac{ 1 }{2}  \ln \left( \frac{1+q  }{ 1+\lambda^t }\right)  
-  \frac{ 1 }{2}  \ln \left( \frac{1-q  }{ 1-\lambda^t }\right)  = \arctanh q -\mu_t
\label{legendremqo}
\end{eqnarray}
with the reciprocal Legendre transformation involves the derivative of Eq. \ref{scgfgalone} with respect to $g$
\begin{eqnarray}
i_t( q)  && = gq-  f_t(g)
\nonumber \\
q && =    \frac{d f_t(g) }{dg} =  \tanh  (g+\mu_t) 
\label{legendremqrecio}
\end{eqnarray}


\section{ Properties of the rate function $I _t^{[m_0]} (m) $ for the intensive magnetization $m$ alone }

\label{app_malone}

The goal of this Appendix is to compute the rate function $I _t^{[m_0]} (m) $ for the intensive magnetization $m$ alone
that was introduced in Eq. \ref{rateImcontraction} of the main text.

\subsection{ Optimization of the joint rate function ${\cal I}_t^{[m_0]}( m,q)$ over the overlap $q$ }

In order to solve the optimization problem of Eq. \ref{rateImcontraction}, 
let us impose the vanishing of the derivative of the joint rate function ${\cal I}_t^{[m_0]}( m,q)$ of Eq. \ref{rateImqderiq} with respect to the overlap $q$
 \begin{eqnarray}
0 && = \partial_q  {\cal I}_t^{[m_0]}( m,q)    =  - \frac{1}{2} \ln \left( \frac{ 1+\lambda^t  }{1-\lambda^t } \right)
+ \frac{ 1 }{4}  \ln \left( \frac{(1+q)^2- (m +  m_0)^2}{(1-q)^2 - (m - m_0)^2}\right) 
\label{rateImqderiqvanish}
\end{eqnarray}
that leads to the second-order equation in $q$
 \begin{eqnarray}
0 && =  (1+\lambda^t)^2 \left[ q^2 -2q+1  - m^2-m_0^2 +2 m m_0 \right]  
  - (1-\lambda^t)^2 \left[q^2 +2q+1 - m^2-m_0^2 +2 m m_0  \right] 
    \nonumber \\
&&=  4 \lambda^t ( q^2 +1  - m^2-m_0^2 )
  -4 (1+\lambda^{2t})  (q - m m_0 )   
\label{secondfull}
\end{eqnarray}
or equivalently after factoring out the global factor $(4 \lambda^t )$
 \begin{eqnarray}
0 && =    q^2 +1  - m^2-m_0^2   - (\lambda^{-t}+\lambda^{t})  (q - m m_0 )   
\nonumber \\
 &&  =  q^2
  -  ( \lambda^t+\lambda^{-t}  ) q
  +   (1-m^2-m_0^2 )
 + ( \lambda^t+\lambda^{-t}  )  m m_0 
\label{second}
\end{eqnarray}
The discriminant 
  \begin{eqnarray}
\Delta && =  ( \lambda^t+\lambda^{-t}  )^2  - 4 (1-m^2-m_0^2 ) - 4 ( \lambda^t+\lambda^{-t}  )  m m_0
 \nonumber \\ 
 && =  ( \lambda^t-\lambda^{-t}  )^2  + 4 \left[ m^2+m_0^2  -  ( \lambda^t+\lambda^{-t}  )  m m_0 \right]
 +  \nonumber \\ 
 && =  ( \lambda^t-\lambda^{-t}  )^2  + 4 ( m-m_0 \lambda^t) (m- m_0\lambda^{-t}  )  
\label{discrimi2}
\end{eqnarray}
leads to the two solutions
 \begin{eqnarray}
q_{\pm} = \frac{ \lambda^t+\lambda^{-t}  \pm \sqrt{\Delta} }{2}
\label{secondpm}
\end{eqnarray}
Since their sum is bigger than 2
 \begin{eqnarray}
q_+ + q_- -2 =  \lambda^t+\lambda^{-t}  - 2 = \left[  \lambda^{\frac{t}{2}} -\lambda^{-\frac{t}{2}}  \right]^2 >0
\label{secondpmsum}
\end{eqnarray}
the solution $q_+$ is bigger than unity $q_+>1$ and is thus outside the interval $q \in [-1,+1]$ of possible overlaps.
The relevant solution is thus $q_-$, and one needs to evaluate 
the rate function $ {\cal I}_t^{[m_0]}( m,q) $ of Eq. \ref{rateImqavecderiusingtypt}
for this special value $q=q_-$ using Eq. \ref{rateImqderim}
 \begin{eqnarray}
  {\cal I}_t^{[m_0]}( m,q_-)  &&  
 =   \frac{m }{4}  \ln \left( \frac{(1+ m)^2 - (  m_0+  q_-)^2}{(1-m)^2 -(m_0-q_-)^2}\right)
 +   \sum_{\epsilon_0=\pm 1}  \frac{1 + \epsilon_0 m_0}{4} 
 \ln \left( \frac{(1+ \epsilon_0 m_0)^2-( m + \epsilon_0 q_-)^2}{(1+ \epsilon_0 m_0)^2( 1- \lambda^{2t})}\right) 
\label{rateImqqmcalcul}
\end{eqnarray}

Using the second-order Eq. \ref{second} satisfied by $q_-$, one obtains the factorizations
 \begin{eqnarray}
(1+q_-)^2 -( m +  m_0)^2 && =( \lambda^t+\lambda^{-t}  +2) (q_- - m m_0) 
= ( \lambda^{\frac{t}{2}}+\lambda^{-\frac{t}{2}} )^2 (q_- - m m_0)
\nonumber \\
(1-q_-)^2 -( m - m_0)^2 && =( \lambda^t+\lambda^{-t}  -2) (q_- - m m_0) 
 = ( \lambda^{\frac{t}{2}}-\lambda^{-\frac{t}{2}} )^2 (q_- - m m_0)
\label{factorssquare}
\end{eqnarray}
that can be used to rewrite the ratio
 \begin{eqnarray}
  \frac{(1+ m)^2 - (  m_0+  q_-)^2}{(1-m)^2 -(m_0-q_-)^2}
&&  = \frac{(1+ m+  m_0+  q_-)(1+ m- m_0-  q_-)}
  {(1-m -m_0+q_-)   (1-m +m_0-q_-)}
  =  \frac{[(1+q_-)^2 -( m +  m_0)^2](1+ m- m_0-  q_-)^2}
  {(1-m -m_0+q_-)^2   [(1-q_-)^2 -( m - m_0)^2]}
  \nonumber \\
&& =   \frac{( \lambda^{\frac{t}{2}}+\lambda^{-\frac{t}{2}} )^2(1+ m- m_0-  q_-)^2}
  {( \lambda^{\frac{t}{2}}-\lambda^{-\frac{t}{2}} )^2(1-m -m_0+q_-)^2  }
  =  \frac{( 1+\lambda^t )^2(1+ m- m_0-  q_-)^2}
  {( 1-\lambda^t )^2(1-m -m_0+q_-)^2  }
\label{essai}
\end{eqnarray}
so that the first contribution of Eq. \ref{rateImqqmcalcul} can be rewritten as
 \begin{eqnarray}
&&  \frac{m }{4}  \ln \left( \frac{(1+ m)^2 - (  m_0+  q_-)^2}{(1-m)^2 -(m_0-q_-)^2}\right)
=   \frac{m }{4}  \ln \left(  \frac{( 1+\lambda^t )^2(1+ m- m_0-  q_-)^2}  {( 1-\lambda^t )^2(1-m -m_0+q_-)^2  }\right)
=   \frac{m }{2}  \ln \left(  \frac{( 1+\lambda^t )(1+ m- m_0-  q_-)}  {( 1-\lambda^t )(1-m -m_0+q_-)  }\right)
\nonumber \\
&& =  \frac{m }{2}  \ln \left(  \frac{ 1+\lambda^t }  { 1-\lambda^t   }\right)
+  \frac{m }{2}  \ln \left(  \frac{1+ m- m_0-  q_-}  {1-m -m_0+q_-  }\right)
\label{FirstContri}
\end{eqnarray}


\subsection{ Final result for the rate function $ I _t^{[m_0]} (m) $ }

In conclusion, the value 
${\mathring q }_t^{[m_0]}(m)$ solving the optimization problem of Eq. \ref{rateImcontraction} 
is given by the solution $q_-$ of Eq. \ref{secondpm}
 \begin{eqnarray}
{\mathring q }_t^{[m_0]}(m) \equiv  \frac{ \lambda^t+\lambda^{-t}  - \sqrt{\Delta} }{2}
=  \frac{ \lambda^t+\lambda^{-t}  - \sqrt{ ( \lambda^t-\lambda^{-t}  )^2  + 4 ( m-m_0 \lambda^t) (m- m_0\lambda^{-t}  )  } }{2}
\label{qmopt}
\end{eqnarray}
and leads to the rate function of Eq. \ref{rateImqqmcalcul} with Eq. \ref{FirstContri}
\begin{eqnarray}
I _t^{[m_0]} (m) =   {\cal I}_t^{[m_0]}( m,{\mathring q }_t^{[m_0]}(m))  
 &&  
 =   \frac{m }{2}  \ln \left(  \frac{ 1+\lambda^t }  { 1-\lambda^t   }\right)
+  \frac{m }{2}  \ln \left(  \frac{1+ m- m_0-  {\mathring q }_t^{[m_0]}(m)}  {1-m -m_0+{\mathring q }_t^{[m_0]}(m)  }\right)
\nonumber \\
&& +   \sum_{\epsilon_0=\pm 1}  \frac{1 + \epsilon_0 m_0}{4} 
 \ln \left( \frac{(1+ \epsilon_0 m_0)^2-( m + \epsilon_0 {\mathring q }_t^{[m_0]}(m))^2}{(1+ \epsilon_0 m_0)^2( 1- \lambda^{2t})}\right) 
\label{rateImcontractionresult}
\end{eqnarray}


\subsection{ Computation of the first derivative $\partial_m I _t^{[m_0]} (m) $ }

Since $q={\mathring q }_t^{[m_0]}(m) $ solves the optimization problem of Eq. \ref{rateImqderiqvanish},
the derivative of the rate function $I _t^{[m_0]} (m) $ of Eq. \ref{rateImcontractionresult} with respect to the magnetization $m$
only involves the partial derivative $\partial_m  {\cal I}_t^{[m_0]}( m,q) $ of Eq. \ref{rateImqderim}
\begin{eqnarray}
\partial_m I _t^{[m_0]} (m) && = \partial_m  {\cal I}_t^{[m_0]}( m,q)  \bigg\vert_{q={\mathring q }_t^{[m_0]}(m)}
+ \left[ \partial_m {\mathring q }_t^{[m_0]}(m)\right]  \partial_q  {\cal I}_t^{[m_0]}( m,q)  \bigg\vert_{q={\mathring q }_t^{[m_0]}(m)}
\nonumber \\ &&
 = \partial_m  {\cal I}_t^{[m_0]}( m,q)  \bigg\vert_{q={\mathring q }_t^{[m_0]}(m)} +0
\nonumber \\ &&  = \frac{1 }{4}  \ln \left( \frac{(1+ m)^2 - (  m_0+  {\mathring q }_t^{[m_0]}(m))^2}{(1-m)^2 -(m_0-{\mathring q }_t^{[m_0]}(m))^2}\right)
\label{rateImcontractionresultderimcalcul}
\end{eqnarray}
that can be rewritten using Eq. \ref{FirstContri}
\begin{eqnarray}
\partial_m I _t^{[m_0]} (m) && =  \frac{m }{2}  \ln \left(  \frac{ 1+\lambda^t }  { 1-\lambda^t   }\right)
+  \frac{m }{2}  \ln \left(  \frac{1+ m- m_0-  {\mathring q }_t^{[m_0]}(m)}  {1-m -m_0+{\mathring q }_t^{[m_0]}(m)  }\right)
\label{rateImcontractionresultderim}
\end{eqnarray}

In particular, when $m$ takes its typical value ${\hat m}=\lambda^t m_0$ of Eq. \ref{bothderivanisheq}, 
the optimal value ${\mathring q }_t^{[m_0]}(m={\hat m}) $ of Eq. \ref{qmopt} coincides with the typical value $\hat q =\lambda^t $ 
of Eq. \ref{bothderivanisheq}
 \begin{eqnarray}
{\mathring q }_t^{[m_0]}({\hat m}=\lambda^t m_0)= \frac{ \lambda^t+\lambda^{-t}  - ( \lambda^{-t} - \lambda^t  ) }{2}
=   \lambda^t = \hat q
\label{qmopttyp}
\end{eqnarray}
with the corresponding vanishing of the rate function of Eq. \ref{rateImcontractionresult}
and of its derivative of Eq. \ref{rateImcontractionresultderim} as expected
\begin{eqnarray}
I _t^{[m_0]} ({\hat m})  &&   =   0
\nonumber \\
\partial_m I _t^{[m_0]} \bigg\vert_{m={\hat m}} && =0
\label{rateImcontractionresultderimtyp}
\end{eqnarray}
while the Gaussian typical fluctuations of order $\frac{1}{\sqrt N}$ of the magnetization $m$ around its typical value ${\hat m}=\lambda^t m_0$ are governed by Eq. \ref{taylorgaussapproxmalone} of the main text.


\end{document}